\documentclass[prl,letterpaper,english,reprint,nofootinbib,aps,superscriptaddress,showpacs,showkeys]{revtex4-1}

\usepackage{babel,calc,amsmath,amsthm,amssymb,graphicx,subfigure,xcolor,comment}
\usepackage{mathdots}
\usepackage[T1]{fontenc}
\usepackage[unicode=true]{hyperref}
\usepackage{booktabs}
\usepackage{threeparttable}

\hypersetup{
     colorlinks=true,       		% false: boxed links; true: colored links
     linkcolor=blue,          	% color of internal links
     citecolor=red,            % color of links to bibliography
     urlcolor=magenta,           	% color of external links
 }
\newtheorem*{theorem*}{Theorem}

\newtheorem*{corollary*}{Corollary}

\newtheorem*{lemma*}{Lemma}

\newtheorem*{proposition*}{Proposition}
\theoremstyle{definition}

\newtheorem*{definition*}{Definition}
\theoremstyle{remark}

\newtheorem*{remark*}{Remark}
\begin{document}
\renewcommand{\figurename}{Fig.}

\title{General Hardy-Type Paradox Based on Bell inequality and its Experimental Test}

\author{Mu Yang}
\affiliation{CAS Key Laboratory of Quantum Information, University of Science and Technology of China, Hefei 230026, People's Republic of China}
\affiliation{CAS Center For Excellence in Quantum Information and Quantum Physics, University of Science and Technology of China, Hefei 230026, People's Republic of China}

\author{Hui-Xian Meng}
\affiliation{Theoretical Physics Division, Chern Institute of Mathematics, Nankai University, Tianjin 300071, People's Republic of China}

\author{Jie Zhou}
\affiliation{Theoretical Physics Division, Chern Institute of Mathematics, Nankai University, Tianjin 300071, People's Republic of China}

\author{Zhen-Peng Xu}
\affiliation{Theoretical Physics Division, Chern Institute of Mathematics, Nankai University, Tianjin 300071, People's Republic of China}

\author{Ya Xiao}
\author{Kai Sun}
\affiliation{CAS Key Laboratory of Quantum Information, University of Science and Technology of China, Hefei 230026, People's Republic of China}
\affiliation{CAS Center For Excellence in Quantum Information and Quantum Physics, University of Science and Technology of China, Hefei 230026, People's Republic of China}

\author{Jing-Ling Chen}
\email{chenjl@nankai.edu.cn}
\affiliation{Theoretical Physics Division, Chern Institute of Mathematics, Nankai University, Tianjin 300071, People's Republic of China}

\author{Jin-Shi Xu}
\email{jsxu@ustc.edu.cn}
\affiliation{CAS Key Laboratory of Quantum Information, University of Science and Technology of China, Hefei 230026, People's Republic of China}
\affiliation{CAS Center For Excellence in Quantum Information and Quantum Physics, University of Science and Technology of China, Hefei 230026, People's Republic of China}

\author{Chuan-Feng Li}
\email{cfli@ustc.edu.cn}
\affiliation{CAS Key Laboratory of Quantum Information, University of Science and Technology of China, Hefei 230026, People's Republic of China}
\affiliation{CAS Center For Excellence in Quantum Information and Quantum Physics, University of Science and Technology of China, Hefei 230026, People's Republic of China}

\author{Guang-Can Guo}
\affiliation{CAS Key Laboratory of Quantum Information, University of Science and Technology of China, Hefei 230026, People's Republic of China}
\affiliation{CAS Center For Excellence in Quantum Information and Quantum Physics, University of Science and Technology of China, Hefei 230026, People's Republic of China}

\date{\today}
\begin{abstract}
Local realistic models cannot completely describe all predictions of quantum mechanics. This is known as Bell's theorem that can be revealed either by violations of Bell inequality, or all-versus-nothing proof of nonlocality. Hardy's paradox is an important all-versus-nothing proof and is considered as ``the simplest form of Bell's theorem''. In this work, we theoretically build the general framework of Hardy-type paradox based on Bell inequality. Previous Hardy's paradoxes have been found to be special cases within the framework. Stronger Hardy-type paradox has been found even for the two-qubit two-setting case, and the corresponding successful probability is about four times larger than the original one, thus  providing a more friendly test for experiment. We also find that GHZ paradox can be viewed as a perfect Hardy-type paradox. Meanwhile, we experimentally test the stronger Hardy-type paradoxes in a two-qubit system. Within the experimental errors, the experimental results coincide with the theoretical predictions.
\end{abstract}

\pacs{03.65.Ud, 03.67.Mn, 42.50.Xa}

	\maketitle

\emph{Introduction.}---In 1935, Einstein, Podolsky and Rosen (EPR) revealed the confliction between quantum mechanics (QM) and local realism by presenting a subtle paradox: either the quantum wave function does not provide a complete description of physical reality, or measuring one particle from a quantum entangled pair instantaneously affects the second particle regardless of how far apart the two entangled particles are~\cite{epr}. Nowadays this paradox is well-known as the famous EPR paradox. The attitude of EPR was incline to consider quantum mechanics as an imcomplete theory by denying the quantum phenomenon of nonlocality. For a long time, the EPR argument has been remained a philosophical debate at the foundation of quantum mechanics until the appearance of Bell's work. In 1964, John Bell discovered Bell's theorem: there is no local realistic theory (with local-hidden-variable (LHV) model) that can reproduce all predictions of quantum mechanics~\cite{bell1964}. To reach his purpose, Bell presented an inequality, which is held for any LHV model, but can be surprisingly violated by quantum entangled states. Bell inequality has provided for the first time a powerful tool to distinguish quantum mechanics essentially from local realistic theory in the Bell-test experiments. Due to which, Bell's theorem has been historically regarded as ``the most profound discovery of science''~\cite{stapp1975}.

In 1969, Clauser, Horne, Shimony and Holt (CHSH) improved the original Bell inequality for two qubits, and they obtained a more experimentally feasible inequality: the CHSH inequality~\cite{CHSH}. Subsequently, Aspect, Grangier and Roger successfully verified Bell nonlocality in experiment~\cite{Aspect81}, despite there were still some locality-loophole and detection-loophole. In 2015, loophole-free Bell-test experiments through violations of the CHSH inequality were carried out to confirm the quantum phenomenon of Bell nonlocality~\cite{loophole-1,loophole-2,loophole-3}. Nowadays, Bell nonlocality has gained significant applications in different quantum information tasks, such as quantum key distribution, communication complexity, quantum information processing and random number generation~\cite{HorodeckiRMP,BrunnerRMP}.

Wondrously, in 1989 Greenberger, Horne and Zeilinger (GHZ) triggered a new approach to study nonlocality: the GHZ theorem~\cite{GHZ89}. Essentially, the GHZ theorem is a kind of \emph{all-versus-nothing} (AVN) proof for Bell nonlocality without using Bell inequality. GHZ theorem is also called GHZ paradox, in which there is a contradiction equality ``$+1=-1$''. Such a sharp contradiction arises if one tries to interpret the quantum result with LHV models, and thus leaves no room for the LHV model to completely describe quantum predictions of the GHZ state. The violation of Bell's inequality demonstrates the incompatibility between the LHV models and quantum mechanics in a statistical manner, while the merit of the AVN proof is just that it may reveal such an incompatibility even with a single run in experiment, therefore needs only fewer experimental measurements.

However, the elegant GHZ argument holds for quantum systems with three and more particles, and unfortunately is not valid for the simplest entangled system: two qubits. To develop the AVN proof (i.e., a logic-contradiction proof) for two qubits, in 1993, Hardy~\cite{hardy93} introduced an alternative paradox:
%, which has been considered as ``the simplest form of Bell's theorem''~\cite{Mermin1995}.

%\begin{eqnarray}\label{eqhardy}
%\left\{
%  \begin{array}{lll}
%    P_1&=&P(A_2 < B_1)=P(A_2=0,B_1=1)=0, \\
%     P_2&=&P(B_1 < A_1)=P(A_1=1,B_1=0)=0, \\
%    P_3&=&P(A_1 < B_2)=P(A_1=0,B_2=1)=0 \\
%  P_4&=&P(A_2 < B_2)=P(A_2=0,B_2=1)\overset{\rm QM} >0. \\
%  \end{array}
%\right.
%\end{eqnarray}
\begin{eqnarray}\label{eqhardy}
\left\{
  \begin{array}{lll}
    P_1&=&P(A_2 < B_1)=0, \\
     P_2&=&P(B_1 < A_1)=0, \\
    P_3&=&P(A_1 < B_2)=0, \\
  P_4&=&P(A_2 < B_2)\overset{\rm QM} >0, \\
  \end{array}
\right.
\end{eqnarray}
where $P(A_i<B_j)$ is the probability for the event $A_i<B_j$, and $P(A_2 < B_2)$ denotes $P(A_2=0, B_2=1)$ for two-qubit, etc. This Hardy paradox consists four probability events. In the classical world, if the first three probability events do not happen, then they will force the fourth probability to be zero. But, according to the different prediction of quantum mechanics, the fourth probability could be greater than zero. The subtle confliction has rendered people to consider Hardy's paradox as ``the simplest form of Bell's theorem''~\cite{Mermin1995}.
The fourth probability is usually called the successful probability, which can reach it maximum $(5\sqrt{5}-11)/2$ ($\approx 9\%$) when one chooses appropriate two-qubit entangled states and projective measurements. Although the successful probability is weak, the two-qubit Hardy paradox has been demonstrated in several  experiments~\cite{Hardy-Exp-1,Hardy-Exp-2,Hardy-Exp-3,Hardy-Exp-4,Hardy-Exp-6,Hardy-Exp-7,Hardy-Exp-8,Hardy-Exp-9}. Later on, Hardy's paradox was theoretically generalized to multi-setting~\cite{Hardy-Exp-3}, multi-partite~\cite{cereceda2004,chen2018} and high-dimensional systems~\cite{chen2003}.

In this Letter, we advance the study of Hardy's paradox. Firstly, we theoretically establish the general framework of Hardy-type paradox based Bell inequality. We find that (i) the previous Hardy paradoxes are merely special cases in the framework; (ii) by extracting some Hardy's constraints from Bell inequality, one can naturally derive some stronger paradoxes with larger successful probabilities, which are more friendly for experimental test; (iii) GHZ paradox can be viewed as a perfect Hardy-type paradox, whose successful probability reaches $100\%$. Secondly, as a physical application, we experimentally test two stronger Hardy-type paradoxes in the two-qubit system, hence demonstrating the sharper confliction between quantum mechanics and local realism.

%As two distinct tools for Bell nonlocality, Bell inequality and the AVN proof indeed have a certain inherent connection. David Mermin is probably the first person to realize such a connection~\cite{Mermin}. By subtracting the forth probability by the first three probabilities, Mermin obtained an inequality $P_4-(P_3+P_2+P_1)\le 0$ (now called the two-qubit Hardy inequality), which is nothing but equivalent to the CHSH inequality. Subsequently, the generalized $N$-qubit Hardy inequality can be derived in the similar way (by subtracting the successful probability by the rest probabilities) based on the generalized $N$-qubit Hardy paradox~\cite{cereceda2004}\cite{chen2018}. Hardy's inequalities are important Bell inequalities, they are particularly significant for proving Gisin's theorem, which states that for any $N$-particle pure states Bell nonlocality and quantum entanglement are equivalent~\cite{yu2012}. It is natural to ask an inverse question: Can one derive some more general Hardy-type paradoxes by starting from Bell inequalities? The answer is positive. The purpose of this Letter is two-folded: (i) We elaborate the theoretical framework for the general Hardy-type paradox based Bell inequality. We find that the previous Hardy paradoxes are only special cases in the framework, and some new Hardy-type paradoxes with higher successful probabilities are also found, which are feasible for experiment. (ii) We experimentally test the stronger Hardy paradoxes in the two-qubit system, thus demonstrating the sharper confliction between quantum mechanics and local realism.

\begin{figure}[t]
	\includegraphics[width=80mm]{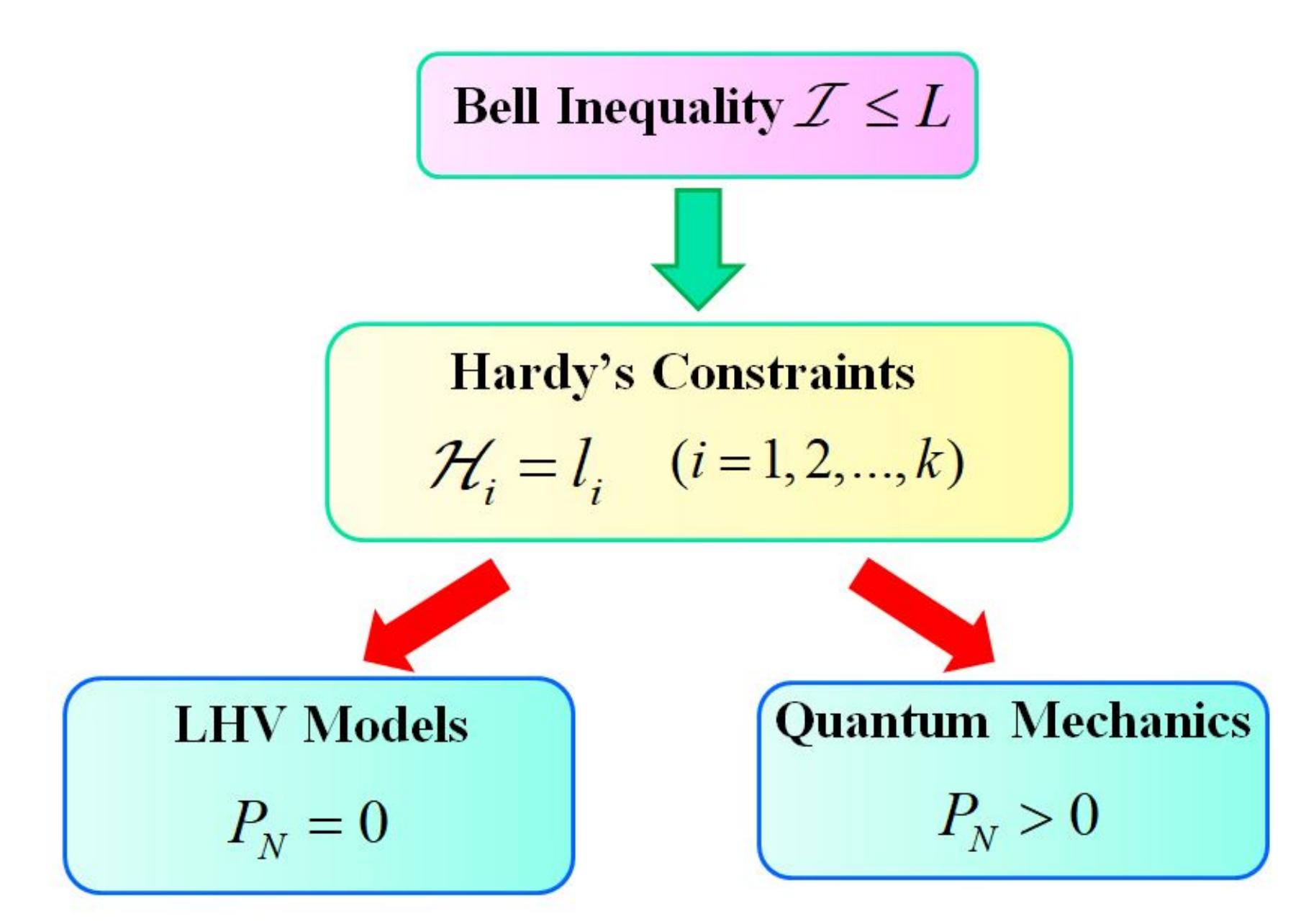}\\
	\caption{(color online). Illustration of the general Hardy-type paradox based on Bell inequality. Here the Bell function $\mathcal{I}=\sum_{i=1}^k \mathcal{H}_i+P_N$. Under the $k$ Hardy's constraints $\mathcal{H}_i=\ell_i$ with $\sum_{i=1}^k \ell_i=L$, the LHV models and quantum mechanics give differently predictions, thus leads to the paradox.
	}\label{fig1}
\end{figure}

\emph{Theoretical Framework.}---%
%Any Bell inequality can be written as $\mathcal{I} \leq C_{\rm LHV}$,
%where $\mathcal{I}$ is the Bell function, and $C_{\rm LHV}$ is the classical bound. In general, Bell function is a linear combination of joint-probabilities as
%$\mathcal{I}=\sum_{j=1}^{N} f_j P_j$ , with $f_j$ being the coefficient (or the weight) of the probability $P_j$.
For any nontrivial Bell inequality, it can be written in the following form
 \begin{eqnarray}\label{bell2}
\mathcal{I} =\sum_{j=1}^{N} f_j P_j \overset{{\rm LHV}}{\leq} L,
 \end{eqnarray}
with $\mathcal{I}$ being the Bell function that is a linear combination of joint-probabilities, $f_j$ the coefficient (or the weight) of the probability $P_j$, and $L$ the classical bound.
%In the determinate LHV model, $P_j$'s are 0 or 1, so the maximal value of Bell function is naturally an integer.
Based on Bell inequality (\ref{bell2}), the general Hardy-type paradox is given as follows:
\begin{eqnarray}\label{ghp}
\left\{
  \begin{array}{lcl}
    \mathcal{H}_1&=& \sum_{j=1}^{i_1} f_j P_j =\ell_1, \\
     \mathcal{H}_2&=& \sum_{j=i_1+1}^{i_2} f_j P_j=\ell_2,   \\
     &\vdots&\\
      \mathcal{H}_{k-1}&=& \sum_{j=i_{k-2}+1}^{i_{k-1}} f_j P_j=\ell_{k-1},\\
      \mathcal{H}_{k}&=& \sum_{j=i_{k-1}+1}^{N-1} f_j P_j=\ell_k,\\
      P_N &\overset{\rm QM} >& 0,
     \end{array}
\right.
\end{eqnarray}
where $\mathcal{H}_i=\ell_i$ ($i=1,2, \cdots, k$) represent Hardy's constraints, $\sum_{i=1}^k \ell_i=L$, and $P_N$ represents the successful probability~\cite{PN}. For the LHV models, because the Bell function is bounded by $L$, thus the $k$ Hardy's constraints must lead to $P_N=0$. But for quantum mechanics, $P_N$ can be greater than zero therefore yields the paradox. The illustration of establishing the general Hardy-type paradox based on Bell inequality can be seen in Fig.~\ref{fig1}.

In the following, we give three concrete examples.

\emph{Example 1.}---The first example is to show that the previous Hardy's paradox can be generated from Bell inequality. The Collins-Gisin-Linden-Massar-Popescu (CGLMP) inequality is a natural extension of the CHSH inequality from two-qubit to two arbitrary $d$-dimensional systems (i.e., two-qudit)~\cite{CGLMP}. The Zohren and Gill's  succinct version~\cite{Gill2008}  of the CGLMP inequality is given by
\begin{eqnarray}\label{gill}
&&P(A_2 < B_2)-P(A_2 < B_1)\nonumber\\
&&\;\;\;-P(B_1 < A_1)-P(A_1 < B_2) \overset{{\rm LHV}}{\leq} 0,
 \end{eqnarray}
Since Bell inequality~(\ref{gill}) is bounded by 0, if one imposes three constraints: $P(A_2 < B_1)=0,\;P(B_1 < A_1)=0,\; P(A_1 < B_2)=0$, then from inequality ~(\ref{gill}), one must have the fourth probability as $P(A_2 < B_2)=0$ for any LHV model. This generates the previous Hardy's paradox for two-qudit, including the two-qubit Hardy's paradox  (\ref{eqhardy}) as a special case. The behavior of the two-qudit Hardy's paradox has been studied in Ref.~\cite{chen2003}, where the successful probability grows up with the increasing dimension $d$. Similarly, one can derive the $N$-qubit Hardy's paradoxes~\cite{cereceda2004} based on the $N$-qubit Hardy's inequalities given in~\cite{chen2018}.

%For $d=2$, it is just the two-qubit Hardy's paradox as in (\ref{eqhardy}), where the successful probability is about $9\%$.

\emph{Example 2.}---The second example is to show that the CHSH inequality, if written in different probability-forms, may generate different Hardy-type paradoxes. In the literature, the CHSH inequality (written in terms of correlation functions) is usually given by $A_1 B_1+A_1 B_2+A_2 B_1-A_2 B_2 \leq 2$. However, when one transforms the correlation-form to the probability-form, actually he can have several different forms of Bell inequality. For instance, the inequality (\ref{gill}) [with $d=2$] is one of the probability-forms. The other form can be the following inequality
\begin{eqnarray}\label{I2222}
\mathcal{I}_{\rm CHSH}&=&P(A_1=1,B_1=1)+P(A_2=1,B_2=0)+\nonumber\\
&&P(A_1=0,B_2=0)+P(A_2=1,B_1=1)+\nonumber\\
&&P(A_1=1,B_2=1)+P(A_2=0,B_1=0)+\nonumber\\
&&P(A_2=0,B_2=1)+P(A_1=0,B_1=0)\nonumber\\
&\overset{\rm LHV}\leq& 3\overset{\rm QM}\leq 2+\sqrt{2}.
\end{eqnarray}
It is worth to mention that the CHSH inequality in the form (\ref{I2222}) is in particular useful. Due to this form, people have successfully proved that Bell nonlocality is tightly bounded by quantum contextually~\cite{CabelloYan1,CabelloYan2}.

Based on the inequality (\ref{I2222}), one can derive the general Hardy-type paradox as
\begin{eqnarray}\label{eqnp0}
\left\{
  \begin{array}{lll}
    \mathcal{H}_1&=&P(A_1=1,B_1=1)+P(A_2=1,B_2=0) \\
    &=&P_1+P_2=1,   \\
    \mathcal{H}_2&=&P(A_1=0,B_2=0)+P(A_2=1,B_1=1) \\
    &=&P_3+P_4=1,   \\
    \mathcal{H}_3&=&P(A_1=1,B_2=1)+P(A_2=0,B_1=0)\\
    &&+P(A_2=0,B_2=1)=P_5+P_6+P_7=1,\\
P_{8}&=&P(A_1=0,B_1=0)>0.
  \end{array}
\right.
\end{eqnarray}
The detail proof is provided in Supplementary Material (SM)~\cite{supp}. Remarkably, the successful probability of the Hardy-type paradox is given by \begin{eqnarray}\label{p8}
P_{8}=P(A_1=0,B_1=0)\approx 0.391179,
\end{eqnarray}
 which is more than four times of the original one ($\approx 9\%$). This paradox not only enhances the sharper confliction between local realism and quantum mechanics, but also is more friendly for the experimental test.

\emph{Example 3.}---The third example is to show that there is the perfect Hardy-type paradox, whose success probability can reach $100\%$. Let us consider the three-qubit Mermin-Ardehali-Belinskii-Klyshko (MABK) inequality~\cite{HorodeckiRMP} that written in the probability-form as
\begin{eqnarray}\label{eqMABK2}
\mathcal{I}_{{\rm{MABK}}}&=&P(A_1+B_2+C_2=0)+P(A_2+B_1+C_2=0)\nonumber\\
&+&P(A_2+B_2+C_1=0)+P(A_1+B_1+C_1=1)\nonumber\\
&&\overset{{\rm{LHV}}}{\leq}3\overset{{\rm{QM}}}{\leq}4.
\end{eqnarray}
Based on which, one can construct the following perfect Hardy-type paradox
\begin{eqnarray}\label{100paradox}
\left\{
  \begin{array}{ll}
    P(A_1+B_2+C_2=0)=1, &  \\
    P(A_2+B_1+C_2=0)=1, &  \\
     P(A_2+B_2+C_1=0)=1, &\\
     P(A_1+B_1+C_1=1)\overset{{\rm{QM}}}{=}1>0.
  \end{array}
\right.
\end{eqnarray}
The connection between GHZ paradox and the perfect Hardy-type paradox is given in SM~\cite{supp}. Actually, GHZ paradox can be viewed as  a kind of Hardy-type paradox, for which the successful probability equals to $100\%$.

%\emph{Remark 1.}---The original Hardy paradox~\cite{hardy93} is two-qubit two-setting paradox, which can be derived directly from CHSH inequality in the form (\ref{gill}) [by taking the dimension as $d=2$], and the maximally successful probability $P_{\rm suc}^{\rm orig}\approx 9\%$. The new Hardy-type paradox (\ref{eqnp0}) is also a two-qubit two-setting paradox, which is derived from an alternative form of the CHSH inequality as in (\ref{I2222}). The proofs of inequality (\ref{gill}) [with $d=2$] and inequality (\ref{I2222}) representing different forms of CHSH inequality is given in SI Sec. IA~\cite{supp}. However, Bell inequality (\ref{I2222}) provides us a more stronger paradox with $P_{\rm suc}^{\rm new}\approx 0.391179$ that is more than four times of the original one. This paradox not only enhances the sharper confliction between local realism and quantum mechanics, but also is more friendly for the experimental test. In the next section, we shall experimentally test the stronger paradox.

\emph{Remark 1.}---In SM~\cite{supp}, we provide more examples for the general Hardy-type paradoxes based on (i) the two-qubit Abner Shimony inequality; and (ii) the two-qubit Bell inequalities realized by the $(2k+1)$-cycle exclusivity graphes. For the case (i), we obtain the stronger paradoxes, for instance, Bell inequality $\mathcal{I}_{4422}\leq 10$ may yield the Hardy-type paradox with successful probability $P_{26}\approx 0.659882$, which is about seven times of the original one ($\approx 9\%$). For the case (ii), we recover the previous ladder Hardy's paradox in~\cite{Hardy-Exp-3}.

%\begin{figure}[t]
%	\begin{center}
%		\includegraphics[width=80mm]{tomo_P8}
%		\caption{(color online). The real part \textbf{a.} and imaginary part \textbf{b.} of the prepared two-qubit state.}	
%		\label{tomo_P8}
%	\end{center}
%\end{figure}

\begin{figure}[t]
	\begin{center}
		\includegraphics[width=80mm]{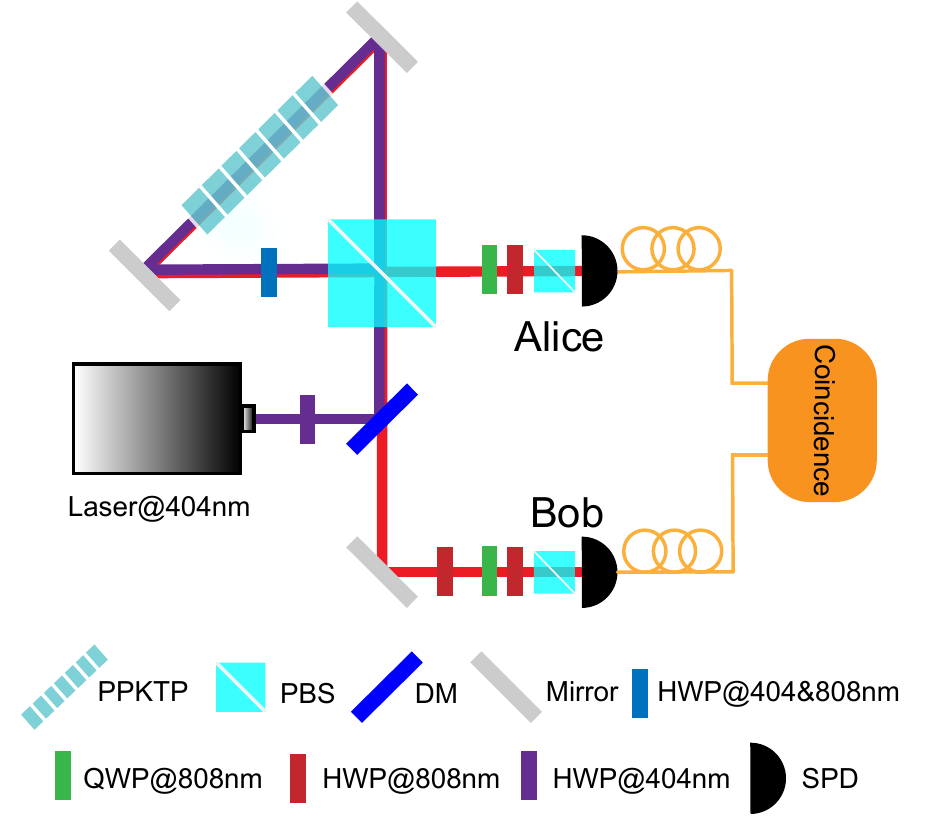}
		\caption{(color online). Experimental setup. An ultraviolet laser with 404 nm are focused to pump a type-II PPKTP crystal located in a Sagnac interferometer to generate polarization-entangled photons. The polarization of the pump laser is set by a half-wave plate (HWP).
		%set before laser is rotated to control the parameter $\theta$.
		The dual-wavelength HWP in the Sagnac interferometer is set to be $45^{\circ}$, which is used to exchange the polarizations of the pump light and down-converted photons. A HWP on Bob's side set in $45^{\circ}$ is used to change the form of entangled state. Quarter-wave plates (QWPs), half-wave plates (HWPs) and polarization beam splitters (PBSs) on both sides of Alice and Bob are used for projective measurements. The photons are detected by single-photon detectors (SPDs), and the signals are sent for coincidence.}
		\label{setup}
	\end{center}
\end{figure}

\emph{Experimental Setup and Results.}---We now experimentally verify the general Hardy-type paradox (\ref{eqnp0}) for the simplest case with two qubits and two measurement settings. The experimental setup is shown in Fig.~\ref{setup}. Polarization-entangle photon pairs are generated through type-II spontaneous parametric down-conversion in a 20 mm-long periodically poled KTP (PPKTP) crystal in a polarization Sagnac interferometer \cite{fedrizzi2007wavelength}. The polarization of the continuous-wave pump laser with a wavelength of 404 nm is rotated to be $\cos\theta\vert H\rangle+\sin\theta\vert V\rangle$ by a half-wave plate (HWP), where $H$ and $V$ represent the horizontal and vertical polarizations, respectively. The pumping light is then divided into two paths by a dual-wavelength polarization beam splitter (PBS) and focused on the crystal. The two photons in the entangled state $\cos\theta|HV\rangle+\sin\theta|VH\rangle$ are sent to Alice and Bob, respectively. The state is then changed to be $|\psi\rangle$= $\cos\theta|HH\rangle+\sin\theta|VV\rangle$ with a HWP set to be $45^{\circ}$ on Bob's side. A quarter-wave plate (QWP), an HWP, and a polarization beam splitter (PBS) on both sides are used for quantum state tomography and projective measurements. To show the maximal confliction of the Hardy paradox, we set $\theta= 0.912$. The fidelity of the prepared state calculated as ${\rm Tr}[\sqrt{\sqrt{\rho_{\rm the}}\rho_{\rm exp}\sqrt{\rho_{\rm the}}}]^2$ is over 0.99, where $\rho_{\rm the}$ and $\rho_{\rm exp}$ represent the theoretical and experimental density matrixes, respectively. The real and imaginary parts of $\rho_{\rm exp}$ are shown in SM~\cite{supp}.

%\begin{table}[htbp]
%	\caption{\label{tab:test} Result of Projective Measurement\tnote{1}}
%	\begin{tabular}{llll}
%		\toprule
%		Alice, Bob  & Coincidence  & Alice, Bob & Coincidence\\
%		Settings & Counts (1s) & Settings & Counts (1s)\\
%		\midrule
%		$HH$ & $3883 (N_{1})$ & $HB_{11}$ & $871 (N_{9})$  \\
%		$HV$ & $8 (N_{2})$ & $VB_{11}$ & $5476 (N_{10})$  \\
%		$VH$ & $11 (N_{3})$ & $HB_{21}$ & $3498 (N_{11})$ \\
%		$VV$ & $6582 (N_{4})$ & $VB_{21}$ & $534 (N_{12})$ \\
%		$A_{11}B_{11}$ & $4765 (N_{5})$ & $A_{11}H$ & $1897 (N_{13})$ \\
%		$A_{21}B_{11}$ & $5137 (N_{6})$ & $A_{11}V$ & $3037 (N_{14})$ \\
%		$A_{11}B_{21}$ & $3783 (N_{7})$ & $A_{21}H$ & $11 (N_{15})$ \\
%		$A_{21}B_{21}$ & $663 (N_{8})$ & $A_{21}V$ & $6513 (N_{16})$ \\
%		\bottomrule
%	\end{tabular}\label{Proj}
%    \begin{tablenotes}
%	   \footnotesize
%\item[1] $A_{i1}, \cos\theta_{ai}\vert H\rangle+\sin\theta_{ai}\vert V\rangle$;
%$B_{i1}, \cos\theta_{bi}\vert H\rangle+\sin\theta_{bi}\vert V\rangle$;\\$\theta_{a1}=0.722,\theta_{a2}=1.504,\theta_{b1}=1.112,\theta_{b2}=0.306$
%    \end{tablenotes}
%\end{table}

\begin{figure}[t]
	\begin{center}
		\includegraphics[width=80mm]{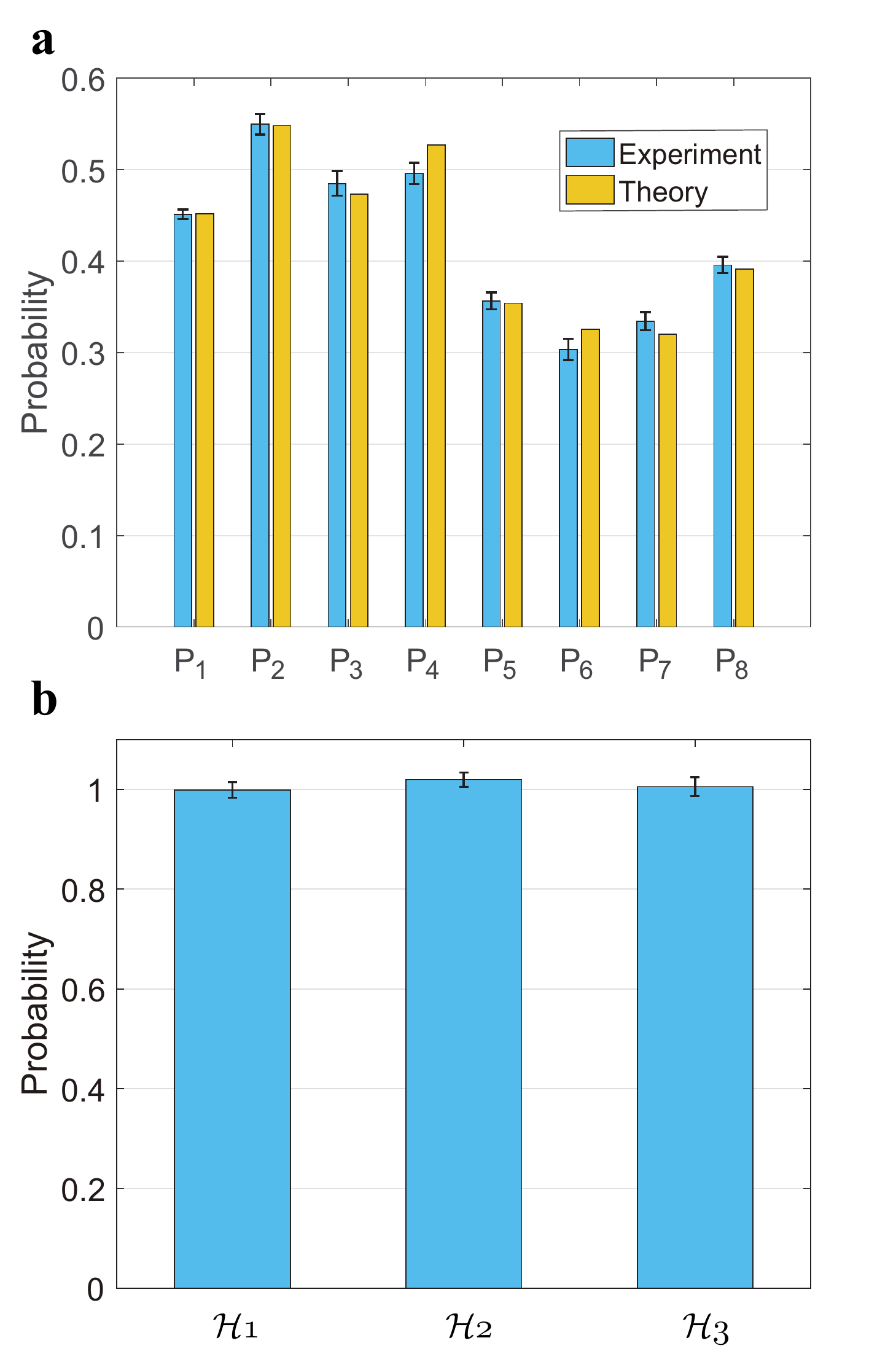}
		\caption{(color online). Experimental results. {\bf a.} The probabilities of $P_{1}$ to $P_{8}$. Blue and yellow columns represent the experimental and theoretical results, respectively. ${\bf b.}$ The experimental values of the three Hardy's constraints given in Eq. (\ref{eqnp0}). The corresponding theoretical predictions are all equal to 1. Error bars are deduced from the counting statistics.}	
		\label{P8}
	\end{center}
\end{figure}

%To measure the probability, We directly measure coincidence counts in 16 different settings, which is shown in Table \ref{Proj}. The coincidence window is 3ns, and the integral time is 1 second.
%We split every probability into a sum of multiple probabilities, and seek the probability of each item separately. More details is shown in SM~\cite{supp}. The $P_{1}-P_{8}$ is expressed as follows:

The eight probabilities of $P_{1}$ to $P_{8}$ as shown in Eq. (\ref{eqnp0}) are then directly detected (see SM~\cite{supp}). To clearly show the confliction, the projective measurement on Alice side $A_{i1}$ ($A_{i}=1$) is chosen to be $\cos\theta_{ai}|H\rangle+\sin\theta_{ai}|V\rangle$ with $\theta_{ai}$=0.722 and 1.504 ($i=1, 2$), respectively. The projective measurement on Bob's side $B_{i1}$ ($B_{i}=1$) is chosen to be $\cos\theta_{bi}|H\rangle+\sin\theta_{bi}|V\rangle$, with $\theta_{bi}$=1.112 and 0.306 ($i=1, 2$). The case of $A_{i0}$ ($A_{i}=0$) and $B_{i0}$ ($B_{i}=0$) can also be obtained since $|A_{i0}\rangle\perp|A_{i1}\rangle$ and $|B_{i0}\rangle\perp|B_{i1}\rangle$~\cite{supp}.
The corresponding probabilities are shown in Fig. \ref{P8}a. The blue columns are the experimental results which agree well with the theoretical predictions represented by yellow columns. $P_{8}=0.3957\pm0.0089$, with about four times larger than the original one \cite{hardy93}. We further show the three Hardy's constraints of $\mathcal{H}_{i}\; (i=1,2,3)$ obtained in experiment in Fig. \ref{P8}b, which are all near equal to 1. All the three constraints are well confirmed within the experimental errors. Error bars are deduced from the counting statistics, which are assumed to be Poissonian distribution. Furthermore, we demonstrated another stronger Hardy-type paradox based on Bell inequality $\mathcal{I}_{4422}\leq 10$, which contains $26$ probabilities measurements. The confliction can reach as high as $P_{26}=0.6802\pm0.0238$ with the confirmation of 9 Hardy's constraints. Experimental results are shown in SM~\cite{supp}.

\emph{Conclusion and Discussion.}---In conclusion, we have advanced the study of Hardy's paradox by building the general theoretical framework of Hardy-type paradox based on Bell inequality. Starting from a certain Bell inequality, one may extract some Hardy's constraints, and then the remain probability will lead to a paradox due to the different predictions of the LHV model and quantum mechanics. Based on the approach, one can recover the previous Hardy's paradoxes and also establish the stronger Hardy-type paradoxes even for two qubits. Meanwhile, as a physical application, we experimentally test the stronger Hardy-type paradox in a two-qubit system, the experimental results coincide with the theoretical predictions within the experimental errors. Moreover, we also find that GHZ paradox can be interestingly viewed as a perfect Hardy-type paradox, thus unifying two apparently different AVN proofs into a single one. In addition, the general Hardy-type paradox is applicable for mixed states, which we shall investigate subsequently.

%As a discussion, there are two interesting observations: (i) For the same Bell inequality  but written in different forms (such as the CHSH inequality written in (\ref{gill}) and (\ref{I2222})), they may result in different Hardy-type paradoxes; (ii) For different Bell inequalities, it may yield the same Hardy-type paradox (for example, the CHSH inequality written in (\ref{gill}) and the 5-cycle Bell inequality generating from exclusivity graph correspond to the same paradox, see SI Sec. IE).

%\begin{acknowledgments}
%\textit{Acknowledgement}.---
This work was supported by the National Key Research and Development Program of China (Grant No. 2016YFA0302700), the National Natural Science Founda-tion of China (Grants No. 61327901, 61725504,  61490711, 11774335 and 11821404), the Key Research Program of Frontier Sciences, Chinese Academy of Sciences (CAS) (Grant No. QYZDY-SSW-SLH003), Anhui Initiative in Quantum Information Technologies (AHY060300 and AHY020100), the Fundamental Research Funds for the Central Universities (Grant No. WK2470000020 and WK2470000026). H.X.M. was supported by the Project funded by China Post-doctoral Science Foundation (No. 2018M631726). J.L.C. was supported by National Natural Science Foundations of China (Grants No.  11475089 and 11875167).

M.Y. and H.X.M. contributed equally to this work.
%\end{acknowledgments}


\begin{thebibliography}{10}
%	\providecommand{\url}[1]{\texttt{#1}}
%	\providecommand{\urlprefix}{URL }
%	\providecommand{\eprint}[2][]{\url{#2}}
\bibitem{epr}
A. Einstein, B. Podolsky, and N. Rosen,
Can Quantum-Mechanical Description of Physical Reality Be Considered Complete?
\href{https://doi.org/10.1103/PhysRev.47.777}{Phys. Rev. \textbf{47}, 777 (1935).}


\bibitem{bell1964}
J. S. Bell,
On the Einstein Podolsky Rosen Paradox.
Physics (Long Island City, N.Y.) \textbf{1}, 195 (1964).

\bibitem{stapp1975}
H. Stapp, Bell's Theorem and World Process, Nuovo Cimento \textbf{29B}, 270 (1975).

\bibitem{CHSH}
J. F. Clauser, M. A. Horne, A. Shimony, and R. A. Holt,
Proposed Experiment to Test Local Hidden-Variable Theories,
\href{https://journals.aps.org/prl/abstract/10.1103/PhysRevLett.23.880}{Phys. Rev. Lett. \textbf{23}, 880 (1969)}.


\bibitem{Aspect81}
A. Aspect, P. Grangier, and G. Roger,
Experimental Tests of Realistic Local Theories via Bell's Theorem,
\href{https://journals.aps.org/prl/abstract/10.1103/PhysRevLett.47.460}{Phys. Rev. Lett. \textbf{47}, 460 (1981)}.

\bibitem{loophole-1}
B. Hensen, \emph{et al.}, Loophole-free Bell inequality violation using electron spins
separated by 1.3 kilometers,
\href{https://www.nature.com/articles/nature15759/}{Nature \textbf{526}, 682 (2015)}.

\bibitem{loophole-2}
M. Giustina, \emph{et al.}, Significant-Loophole-Free Test of Bell's Theorem with
Entangled Photons, \href{https://doi.org/10.1103/PhysRevLett.115.250401}{Phys. Rev. Lett. \textbf{115}, 250401 (2015)}.

\bibitem{loophole-3}
L. Shalm, \emph{et al.}, Strong Loophole-Free Test of Local Realism,  \href{https://doi.org/10.1103/PhysRevLett.115.250402}{Phys. Rev. Lett. \textbf{115}, 250402 (2015)}.

\bibitem{HorodeckiRMP}
R. Horodecki, P. Horodecki, M. Horodecki, K. Horodecki,
Quantum entanglement,
\href{https://journals.aps.org/rmp/abstract/10.1103/RevModPhys.81.865}{Rev. Mod. Phys. \textbf{81}, 865 (2009)}.

\bibitem{BrunnerRMP}
N. Brunner, D. Cavalcanti, S. Pironio, V. Scarani, and S. Wehner,
Bell nonlocality,
\href{https://journals.aps.org/rmp/abstract/10.1103/RevModPhys.86.419}{Rev. Mod. Phys. \textbf{86}, 419 (2014)}.

\bibitem{GHZ89}
D. M. Greenberger, M. A. Horne, and A. Zeilinger,
In \emph{Bell's Theorem, Quantum Theory, and Conceptions of the Universe}, edited by
M. Kafatos, (Kluwer, Dordrecht, 1989) p. 69.

\bibitem{hardy93}
L. Hardy,
Nonlocality for two particles without inequalities for almost all entangled states.
\href{https://doi.org/10.1103/PhysRevLett.71.1665}{Phys. Rev. Lett. \textbf{71}, 1665 (1993).}


\bibitem{Mermin1995}
N. D. Mermin, in Fundamental Problems in Quantum Theory,
edited by D. M. Greenberger and A. Zeilinger, Special issue of
\href{https://nyaspubs.onlinelibrary.wiley.com/doi/abs/10.1111/j.1749-6632.1995.tb39001.x}{Ann. N. Y. Acad. Sci. 755, 616 (1995)}.


%\bibitem{kar97}
%G. Kar,
%Testing Hardy's nonlocality with $n$ spin-1/2 particles.
%\href{https://doi.org/10.1103/PhysRevA.56.1023}{Phys. Rev. A \textbf{56}, 1023 (1997).}

\bibitem{Hardy-Exp-1}
J. R. Torgerson, D. Branning, C. H. Monken, and L. Mandel,
Experimental demonstration of the violation of local realism without Bell inequalities.
\href{https://doi.org/10.1016/0375-9601(95)00486-M}{Phys. Lett. A \textbf{204}, 323 (1995).}

\bibitem{Hardy-Exp-2}
G. Di Giuseppe, F. De Martini, and D. Boschi,
Experimental test of the violation of local realism in quantum mechanics without Bell inequalities.
\href{https://doi.org/10.1103/PhysRevA.56.176}{Phys. Rev. A \textbf{56}, 176 (1997).}

\bibitem{Hardy-Exp-3}
D. Boschi, S. Branca, F. De Martini, and L. Hardy,
Ladder Proof of Nonlocality without Inequalities: Theoretical and Experimental Results.
\href{https://doi.org/10.1103/PhysRevLett.79.2755}{Phys. Rev. Lett. \textbf{79}, 2755 (1997).}

\bibitem{Hardy-Exp-4}
M. Barbieri, C. Cinelli, F. De Martini, and P. Mataloni,
Test of quantum nonlocality by full collection of polarization entangled photon pairs.
\href{https://doi.org/10.1140/epjd/e2004-00199-6}{Eur. Phys. J. D \textbf{32}, 261 (2005).}

%\bibitem{Hardy-Exp-5}
%J. S. Lundeen and A. M. Steinberg,
%Experimental Joint Weak Measurement on a Photon Pair as a Probe of Hardy's Paradox.
%\href{https://doi.org/10.1103/PhysRevLett.102.020404}{Phys. Rev. Lett. \textbf{102}, 020404 (2009).}

\bibitem{Hardy-Exp-6}
G. Vallone, I. Gianani, E. B. Inostroza, C. Saavedra, G. Lima, A. Cabello, and P. Mataloni,
Testing Hardy's nonlocality proof with genuine energy-time entanglement.
\href{https://doi.org/10.1103/PhysRevA.83.042105}{Phys. Rev. A \textbf{83}, 042105 (2011).}

\bibitem{Hardy-Exp-7}
A. Fedrizzi, M. P. Almeida, M. A. Broome, A. G. White, and M. Barbieri,
Hardy's Paradox and Violation of a State-Independent Bell Inequality in Time.
\href{https://doi.org/10.1103/PhysRevLett.106.200402}{Phys. Rev. Lett. \textbf{106}, 200402 (2011).}

\bibitem{Hardy-Exp-8}
L. Chen and J. Romero,
Hardy' nonlocality proof using twisted photons.
\href{https://doi.org/10.1364/OE.20.021687}{Opt. Express \textbf{20}, 21687 (2012).}

\bibitem{Hardy-Exp-9}
E. Karimi, F. Cardano, M. Maffei, C. de Lisio, L. Marrucci, R. W. Boyd, and E. Santamato,
Hardy's paradox tested in the spin-orbit Hilbert space of single photons.
\href{https://doi.org/10.1103/PhysRevA.89.032122}{Phys. Rev. A \textbf{89}, 032122 (2014).}

\bibitem{cereceda2004}
J. L. Cereceda,
Hardy's nonlocality for generalized n-particle ghz states.
\href{https://doi.org/10.1016/j.physleta.2004.06.004}{Phys. Lett. A \textbf{327}, 433 (2004).}

\bibitem{chen2018}
S. H. Jiang, Z. P. Xu, H. Y. Su, A. K. Pati, and J. L. Chen, Generalized
Hardy's Paradox. \href{https://journals.aps.org/prl/abstract/10.1103/PhysRevLett.120.050403}{Phys. Rev. Lett. \textbf{120}, 050403 (2018)}.


\bibitem{chen2003}
J. L. Chen, A. Cabello, Z. P. Xu, H. Y. Su, C. Wu, and L. C. Kwek,
Hardy's paradox for high-dimensional systems.
\href{https://doi.org/10.1103/PhysRevA.88.062116}{Phys. Rev. A \textbf{88}, 062116 (2013).}


\bibitem{PN}
$P_N$ may contain several probabilities when one considers the high-dimensional systems or multipartite systems, for instance, in Hardy's paradox  (\ref{eqhardy}),  $P(A_2 < B_2)=\sum_{b>a=0}^{d-1}P(A_2=a, B_2=b)$ may contain several terms of probability if one considers the two-qudit case.


\bibitem{CGLMP}
D. Collins, N. Gisin, N. Linden, S. Massar, and S. Popescu,
Bell Inequalities for Arbitrarily High-Dimensional Systems,
\href{https://journals.aps.org/prl/abstract/10.1103/PhysRevLett.88.040404}{Phys. Rev. Lett. \textbf{88}, 040404 (2002)}.


\bibitem{Gill2008}
S. Zohren and R. D. Gill,
Maximal Violation of the Collins-Gisin-Linden-Massar-Popescu Inequality for Infinite Dimensional States.
\href{https://doi.org/10.1103/PhysRevLett.100.120406}{Phys. Rev. Lett. \textbf{100}, 120406 (2008).}

%\bibitem{bellinteger}
%It is not interesting to consider the coefficients $f_j$ as  irrational numbers, because tight Bell inequalities have not been encountered for this case in the literature. It is sufficient to consider $f_j$ as integers, because when $f_j$ are some rational fractions, one can always multiply the largest denominator to the both sides of Bell inequality, such that the coefficients in the Bell function then become integers. Furthermore, it is enough to consider $f_j$ as positive integers in any Bell inequality with the help of the normalization condition of joint-probabilities. For instance, by considering $P(A_2 < B_1)=1-P(A_2 \geq B_1)$ etc, inequality (\ref{gill}) becomes
%$P(A_2 < B_2)+P(A_2 \geq B_1)+P(B_1 \geq A_1)+P(A_1 \geq B_2) {\leq} 3$,
%for which coefficients $f_j$'s are all positive integers, and the three Hardy's constraints now turn to $P(A_2 \geq B_1)=P(B_1 \geq A_1)=P(A_1 \geq B_2)=1$.




\bibitem{CabelloYan1}
 A. Cabello, Simple Explanation of the Quantum Violation of a Fundamental Inequality, \href{https://journals.aps.org/prl/abstract/10.1103/PhysRevLett.110.060402}{Phys. Rev. Lett. \textbf{110}, 060402 (2013)}.

\bibitem{CabelloYan2}
B. Yan, Quantum Correlations are Tightly Bound by the Exclusivity Principle, \href{https://journals.aps.org/prl/abstract/10.1103/PhysRevLett.110.260406}{Phys. Rev. Lett. \textbf{110}, 260406 (2013)}.

\bibitem{supp}
See Supplementary Material for more details.

\bibitem{fedrizzi2007wavelength} A. Fedrizzi, T. Herbst, A. Poppe, T. Jennewein, and A. Zeilinger,
A wavelength-tunable fiber-coupled source of narrowband entangled photons,
\href{https://www.osapublishing.org/DirectPDFAccess/BBEC47B3-AD3F-E1BC-2A222200404CD2E5_144702/oe-15-23-15377.pdf?da=1&id=144702&seq=0&mobile=no}{Opt. Express \textbf{15}, 15377 (2007).}



%\bibitem{Mermin}
%N. D. Mermin, What's wrong with the tempatation? \href{https://physicstoday.scitation.org/doi/10.1063/1.2808514}{Phys. Today \textbf{47}(6), 9 (1994)}; What's wrong with the punctation? \href{https://physicstoday.scitation.org/doi/10.1063/1.2808722}{Phys. Today \textbf{47}(11), 119 (1994)}.
%
%\bibitem{yu2012}
%S. Yu, Q. Chen, C. Zhang, C. H. Lai, and C. H. Oh, All Entangled Pure States
%Violate a Single Bell's Inequality, \href{https://journals.aps.org/prl/abstract/10.1103/PhysRevLett.109.120402}{Phys. Rev. Lett. \textbf{109}, 120402 (2012)}.
%

\end{thebibliography}
\end{document}

%\title{General Hardy-Type Paradox Based on Bell inequality and its Experimental Test}

\title{Supplemental Material for ``General Hardy-Type Paradox Based on Bell inequality and its Experimental Test''}

\author{Mu Yang}
\affiliation{CAS Key Laboratory of Quantum Information, University of Science and Technology of China, Hefei 230026, People's Republic of China}
\affiliation{CAS Center For Excellence in Quantum Information and Quantum Physics,, University of Science and Technology of China, Hefei 230026, People's Republic of China}

\author{Hui-Xian Meng}
\affiliation{Theoretical Physics Division, Chern Institute of Mathematics, Nankai University, Tianjin 300071, People's Republic of China}

\author{Jie Zhou}
\affiliation{Theoretical Physics Division, Chern Institute of Mathematics, Nankai University, Tianjin 300071, People's Republic of China}

\author{Zhen-Peng Xu}
\affiliation{Theoretical Physics Division, Chern Institute of Mathematics, Nankai University, Tianjin 300071, People's Republic of China}

\author{Ya Xiao}
\author{Kai Sun}
\affiliation{CAS Key Laboratory of Quantum Information, University of Science and Technology of China, Hefei 230026, People's Republic of China}
\affiliation{CAS Center For Excellence in Quantum Information and Quantum Physics, University of Science and Technology of China, Hefei 230026, People's Republic of China}

\author{Jing-Ling Chen}
%\email{chenjl@nankai.edu.cn}
\affiliation{Theoretical Physics Division, Chern Institute of Mathematics, Nankai University, Tianjin 300071, People's Republic of China}

\author{Jin-Shi Xu}
%\email{jsxu@ustc.edu.cn}
\affiliation{CAS Key Laboratory of Quantum Information, University of Science and Technology of China, Hefei 230026, People's Republic of China}
\affiliation{CAS Center For Excellence in Quantum Information and Quantum Physics, University of Science and Technology of China, Hefei 230026, People's Republic of China}

\author{Chuan-Feng Li}
%\email{cfli@ustc.edu.cn}
\affiliation{CAS Key Laboratory of Quantum Information, University of Science and Technology of China, Hefei 230026, People's Republic of China}
\affiliation{CAS Center For Excellence in Quantum Information and Quantum Physics, University of Science and Technology of China, Hefei 230026, People's Republic of China}

\author{Guang-Can Guo}
\affiliation{CAS Key Laboratory of Quantum Information, University of Science and Technology of China, Hefei 230026, People's Republic of China}
\affiliation{CAS Center For Excellence in Quantum Information and Quantum Physics, University of Science and Technology of China, Hefei 230026, People's Republic of China}

\date{\today}

%\pacs{03.65.Ud, 03.67.Mn, 42.50.Xa}

\maketitle

\section{The theoretical part}

\subsection{The CHSH inequality written in different probability-forms}

The Clauser-Horne-Shimony-Holt (CHSH) inequality \cite{chsh} is a well-known Bell inequality for two qubits. If one writes the CHSH  inequality in terms of correlation functions, then he obtais
\begin{align}\label{CHSH}
\langle A_1B_1\rangle+\langle A_1B_2\rangle+\langle A_2B_1\rangle-\langle A_2B_2\rangle\overset{{\rm{LHV}}}{\leq}2,
%\overset{{\rm{QM}}}{\leq}2\sqrt{2},
\end{align}
where $A_i,B_j=\pm1$ ($i,j=1,2$), and $\langle A_iB_j\rangle$  are the correlation functions ($A_i$ measured on Alice's particle and $B_j$ measured on Bob's particle).

For a two-qubit system, there is a relation between correlation function and joint-probabilities:
\begin{eqnarray}\label{qij}
Q_{ij}&=&\langle A_i B_j\rangle=\sum_{a,b=0}^1 (-1)^{a+b}\; P(A_i=a,B_j=b)\nonumber\\
&=& P(A_i=0,B_j=0)+P(A_i=1,B_j=1)-P(A_i=0,B_j=1)-P(A_i=1,B_j=0)\nonumber\\
&=& 2[P(A_i=0,B_j=0)+P(A_i=1,B_j=1)]-1\nonumber\\
&=& 1-2[P(A_i=0,B_j=1)+P(A_i=1,B_j=0)].
 \end{eqnarray}
or
\begin{eqnarray}\label{qij-1}
Q_{ij}&=&\langle A_i B_j\rangle= P(A_i =B_j)-P(A_i\neq B_j)\nonumber\\
&=& 2P(A_i=B_j)-1\nonumber\\
&=& 1-2P(A_i\neq B_j).
 \end{eqnarray}
By substituting
\begin{eqnarray}\label{qij-2}
\langle A_1 B_1\rangle&=& 2[P(A_1=0,B_1=0)+P(A_1=1,B_1=1)]-1,\nonumber\\
\langle A_1 B_2\rangle&=& 2[P(A_1=0,B_2=0)+P(A_1=1,B_2=1)]-1,\nonumber\\
\langle A_2 B_1\rangle&=& 2[P(A_2=0,B_1=0)+P(A_1=1,B_1=1)]-1,\nonumber\\
\langle A_2 B_2\rangle&=& 1-2[P(A_2=0,B_2=1)+P(A_2=1,B_2=0)],
 \end{eqnarray}
into (\ref{CHSH}), one immediately has the CHSH inequality written in terms of joint-probabilities as
\begin{eqnarray}\label{eqsp}
\begin{split}
\mathcal{I}_{{\rm{CHSH}}}=\sum_{i=1}^8P_i&=P(A_1=1,B_1=1)+P(A_2=1,B_2=0)+P(A_1=0,B_2=0)+P(A_2=1,B_1=1)\\
&~~~~+P(A_1=1,B_2=1)+P(A_2=0,B_1=0)+P(A_2=0,B_2=1)+P(A_1=0,B_1=0)\\
&\overset{{\rm LHV}}\leq 3,
\end{split}
\end{eqnarray}
which is just the inequality (5) in the main text.

However, when one writes the CHSH inequality in terms of joint-probabilties, actually the probability-form is not unique. For instance, let us consider
the Zohren and Gill's  succinct version of the CGLMP inequality for two-qudit~\cite{CGLMP}\cite{Gill2008}\cite{chen2003}, which is given as follows:
\begin{eqnarray}\label{gill}
&&P(A_2 < B_2)-P(A_2 < B_1)-P(B_1 < A_1)-P(A_1 < B_2) \overset{{\rm LHV}}{\leq} 0,
 \end{eqnarray}
where $P(A_i<B_j)$ is the probability of $A_i<B_j$. For two-qubit, the inequality becomes
\begin{eqnarray}\label{gill-1}
&&P(A_2=0, B_2=1)-P(A_2=0, B_1=1)-P(A_1=1, B_1=0)-P(A_1=0, B_2=1) \overset{{\rm LHV}}{\leq} 0,
 \end{eqnarray}
which is also equivalent to the CHSH inequality (\ref{CHSH}).

{\it Proof.}--- Due to the following relation between joint-probabilities and correlation functions
\begin{eqnarray}\label{pc}
&& P(A_i=a,B_j=b)=\left\langle\frac{I_2+(-1)^{a}A_i}{2}\otimes\frac{I_2+(-1)^{b}B_j}{2}\right\rangle,\;\;\;\; a,b\in\{0,1\},
\end{eqnarray}
with $I_2$ the $2\times 2$ unit matrix, then the left-hand side of the
inequality (\ref{gill-1}) can be reduced to
\begin{eqnarray}\label{gill-2}
\begin{split}
&~~~~P(A_2 < B_2)-P(A_2 < B_1)-P(B_1 < A_1)-P(A_1 < B_2)\\
&=P(A_2=0,B_2=1)-P(A_2=0,B_1=1)-P(A_1=1,B_1=0)-P(A_1=0, B_2=1)\\
&=\left\langle\frac{I_2+A_2}{2}\otimes\frac{I_2-B_2}{2}\right\rangle-
\left\langle\frac{I_2+A_2}{2}\otimes\frac{I_2-B_1}{2}\right\rangle
-\left\langle\frac{I_2-A_1}{2}\otimes\frac{I_2+B_1}{2}\right\rangle
-\left\langle\frac{I_2+A_1}{2}\otimes\frac{I_2-B_2}{2}\right\rangle\\
&=-\frac{1}{2}-\frac{1}{4}(\langle A_2B_2\rangle-\langle A_2B_1\rangle-\langle A_1B_1\rangle-\langle A_1B_2\rangle)\overset{{\rm LHV}}{\leq} 0,
\end{split}
 \end{eqnarray}
which may yield the inequality (\ref{CHSH}). Thus, the inequality (\ref{gill-1}) is also one of probability-forms for the CHSH inequality. This ends the proof.

In summary, inequalities (\ref{eqsp}) and (\ref{gill-1}) are both the CHSH inequality but written in different probability-forms.

\subsection{General Hardy-type paradoxes based on the CHSH inequality in two different probability-forms}

{\it Case 1.}---For the CHSH inequality written in the form (\ref{gill-1}), one can have Hardy's paradox as
\begin{equation}\label{eqH}
\left\{
     \begin{array}{ll}
       P(A_2<B_{1})=0, \\
       P(B_{1}<A_{1})=0, \\
      P(A_1<B_2)=0,\\
       P(A_2<B_2)\overset{{\rm{QM}}}=\frac{5\sqrt{5}-11}{2}>0,
     \end{array}
   \right.
 \end{equation}
This paradox is equivalent to the original Hardy's paradox~\cite{hardy93}, for which the maximal successful probability equals to $(5\sqrt{5}-11)/2$($\approx 9\%$).

The following is the proof. Assume that the two-qubit system is in the sate $|\psi\rangle$, and the local measurements $A_{i}$ and $B_j$ $(i,j=1,2)$ can be represented by
\begin{eqnarray}
A_i&=& |A_{i0}\rangle\langle A_{i0}|-|A_{i1}\rangle\langle A_{i1}|,\nonumber\\
B_j&=& |B_{j0}\rangle\langle B_{j0}|-|B_{j1}\rangle\langle B_{j1}|,
\end{eqnarray}
 where
 $\{|A_{i0}\rangle,|A_{i1}\rangle\}$ and $\{|B_{j0}\rangle,|B_{j1}\rangle\}$ are the orthonormal basis, with $|A_{i0}\rangle\langle A_{i0}|+|A_{i1}\rangle\langle A_{i1}|=I_2$ and
 $|B_{j0}\rangle\langle B_{j0}|+|B_{j1}\rangle\langle B_{j1}|=I_2$.  Then we obtain
\begin{eqnarray}
&&P(A_i=0,B_j=0)=|\langle\psi|A_{i0}\rangle|B_{j0}\rangle|^2, \;\;P(A_i=0,B_j=1)=
|\langle\psi|A_{i0}\rangle|B_{j1}\rangle|^2,\nonumber\\
&&P(A_i=1,B_j=0)=|\langle\psi|A_{i1}\rangle|B_{j0}\rangle|^2,\;\; P(A_i=1,B_j=1)=
|\langle\psi|A_{i1}\rangle|B_{j1}\rangle|^2.
\end{eqnarray}
By taking
\begin{eqnarray}
&&|\psi\rangle=\alpha|00\rangle+\beta|11\rangle,|A_{20}\rangle=\left(
                              \begin{array}{c}
                                x\\
                                y \\
                              \end{array}
                            \right),
                           |A_{21}\rangle\bot |A_{20}\rangle,\nonumber\\
&&|A_{10}\rangle=\frac{1}{\sqrt{\alpha^{4}x^2+\beta^{4}y^2}}\left(
                              \begin{array}{c}
                                \alpha^{2}x \\
                                \beta^{2}y \\
                              \end{array}
                            \right),|A_{11}\rangle\bot |A_{10}\rangle,\;\;\;|B_{10}\rangle=\frac{1}{\sqrt{\alpha^{2}x^2+\beta^{2}y^2}}\left(
                              \begin{array}{c}
                                 \alpha x \\
                                \beta y \\
                              \end{array}
                            \right),|B_{11}\rangle\bot |B_{10}\rangle,\nonumber\\
&&|B_{20}\rangle=\frac{1}{\sqrt{\alpha^{6}x^2+\beta^{6}y^2}}\left(
                              \begin{array}{c}
                                 \alpha^{3}x \\
                                \beta^{3}y \\
                              \end{array}
                            \right),|B_{21}\rangle =\frac{1}{\sqrt{\alpha^{6}x^2+\beta^{6}y^2}}\left(
                              \begin{array}{c}
                                 \beta^{3}y \\
                                -\alpha^{3}x\\
                              \end{array}
                            \right),
\end{eqnarray}
we then have
\begin{eqnarray}
\label{eqHprime1}
\left\{
     \begin{array}{ll}
       P(A_2<B_{1})=P(A_2=0,B_1=1)=0, \\
       P(B_{1}<A_{1})=P(A_1=1,B_1=0)=0, \\
      P(A_1<B_2)=P(A_1=0,B_2=1)=0,\\
     \end{array}
   \right.
\end{eqnarray}
and
\begin{eqnarray}
\label{eqHprime2}
       P(A_2<B_2)=P(A_2=0,B_2=1)=\frac{(\alpha\beta^{3}-\beta \alpha^3)^2\;\; x^2 y^2}
{x^2 \alpha^{6} +y^2 \beta^{6}}.
\end{eqnarray}

For the parameters
\begin{eqnarray}
&&\alpha=\sqrt{\frac{1}{2}\left(1 + \sqrt{-13 + 6 \sqrt{5}}\right)},\;\; \beta=\sqrt{\frac{1}{2}\left(1-\sqrt{-13 + 6 \sqrt{5}}\right)},
\end{eqnarray}
\begin{eqnarray}
&&x=\frac{\beta^3}{\alpha^3+\beta^3}=\sqrt{\frac{1}{22}\left(11 - \sqrt{55(-3 + 2\sqrt{5})}\right)},\;\;y=1-x^2=\sqrt{\frac{1}{22}\left(11 + \sqrt{55(-3 + 2\sqrt{5})}\right)}.
\end{eqnarray}
then the three constraint conditions in (\ref{eqH}) are satisfied (or see Eq. (\ref{eqHprime1})) and the successful probability attains its maximum value as
\begin{eqnarray}
&&\max\{P(A_2<B_2)\}=\frac{5\sqrt{5}-11}{2}\approx 9\%.
\end{eqnarray}

{\it Case 2.}---For the CHSH inequality written in the form (\ref{eqsp}), one can have the following general Hardy-type paradox

 %{\it Example 1}
% \begin{itemize}
% \item $k$=1: \begin{eqnarray}\label{eqnp01}
%\left\{
%  \begin{array}{ll}
%    \mathcal{H}_1=I_{{\rm{CHSH}}}-P(A_1=0,B_1=0)=\sum_{i=1}^7P_i=3, &  \\
%P_8=P(A_1=0,B_1=0)\overset{{\rm{QM}}}\approx 0.41399>0,
%  \end{array}
%\right.
%\end{eqnarray}
%is a general Hardy-type paradox;
%  \item $k$=2: \begin{eqnarray}\label{eqnp02}
%\left\{
%  \begin{array}{ll}
%    \mathcal{H}_1=P(A_1=1,B_2=1)+P(A_2=1,B_2=0)=P_1+P_2=1, &  \\
%    \mathcal{H}_2=I_{{\rm{CHSH}}}-P(A_1=0,B_1=0)-\mathcal{H}_1=\sum_{i=3}^7P_i=2, &  \\
%P_8=P(A_1=0,B_1=0)\overset{{\rm{QM}}}\approx 0.404638>0, &
%  \end{array}
%\right.
%\end{eqnarray}
%is a general Hardy-type paradox;
% \item $k$=3:
\begin{eqnarray}\label{eqnp03}
\left\{
  \begin{array}{ll}
    \mathcal{H}_1=P(A_1=1,B_1=1)+P(A_2=1,B_2=0)=P_1+P_2=1, &  \\
    \mathcal{H}_2=P(A_1=0,B_2=0)+P(A_2=1,B_1=1)=P_3+P_4=1, &  \\
    \mathcal{H}_3=P(A_1=1,B_2=1)+P(A_2=0,B_1=0)+P(A_2=0,B_2=1)=P_5+P_6+P_7=1, &\\
P_{8}=P(A_1=0,B_1=0)\overset{{\rm{QM}}}\approx 0.391179>0, &
  \end{array}
\right.
\end{eqnarray}

%\end{itemize}

The following is the proof. In local realistic theory, suppose the output of measurement $A_1$ were $0$, then by the first Hardy's constraint in (\ref{eqnp03}) one has that the outputs of measurements $A_2$ and $B_2$ are $1$ and $0$ respectively, which contradict with the third Hardy's constraint in (\ref{eqnp03}). Thus, the output of measurement $A_1$ must be $1$, which implies that the outputs of both measurements $A_2$ and $B_1$ are $1$ by the second Hardy's constraint. Hence,  the output of measurement $B_2$ is also $1$ by the third constraint condition in (\ref{eqnp03}). This means that by requiring $\mathcal{H}_i=1$ ($i=1,2,3$), one obtains $A_1=B_1=A_2=B_2=1$. Thus, for local correlations satisfying (\ref{eqnp03}), $P_{8}=P(A_1=0,B_1=0)=0$ holds.

In quantum mechanics, the two-qubit pure state $|\psi\rangle$ and local measurements $\{A_1,A_2\}$ and $\{B_1,B_2\}$ can be represented by
\begin{eqnarray}\label{eqmAB}
\left\{
  \begin{array}{ll}
|\psi\rangle=(\cos{\theta},0,0,\sin{\theta})^T, &  \\
    A_i=|A_{i0}\rangle\langle A_{i0}|-|A_{i1}\rangle\langle A_{i1}|, &  \\
    B_j=|B_{j0}\rangle\langle B_{j0}|-|B_{j1}\rangle\langle B_{j1}|, & \\
   |A_{i1}\rangle=(\cos{\theta_{ai}},\sin{\theta}_{ai})^T, \;\;\;|A_{i0}\rangle=(\sin{\theta}_{ai},-\cos{\theta_{ai}})^T,&  \\
    |B_{j1}\rangle=(\cos{\theta_{bj}},\sin{\theta}_{bj})^T,\;\;\;|B_{j0}\rangle=(\sin{\theta}_{bj},-\cos{\theta_{bj}})^T, & \\
  \end{array}
\right.
\end{eqnarray}
for $i,j=1, 2.$ Then we obtain
\begin{eqnarray}
\left\{
  \begin{array}{ll}
    P(A_i=0,B_j=0)=|\langle\psi|A_{i0}\rangle|B_{j0}\rangle|^2=(\cos{\theta}\sin{\theta_{ai}}\sin{\theta_{bj}}
+\sin{\theta}\cos{\theta}_{ai}\cos{\theta}_{bj})^2, \\
    P(A_i=0,B_j=1)=
|\langle\psi|A_{i0}\rangle|B_{j1}\rangle|^2=(\cos{\theta}\sin{\theta_{ai}}\cos{\theta}_{bj}
-\sin{\theta}\cos{\theta}_{ai}\sin{\theta_{bj}})^2,\\
    P(A_i=1,B_j=0)=|\langle\psi|A_{i1}\rangle|B_{j0}\rangle|^2=(\cos{\theta}\cos{\theta}_{ai}\sin{\theta_{bj}}
-\sin{\theta}\sin{\theta_{ai}}\cos{\theta}_{bj})^2, \\
    P(A_i=1,B_j=1)=
|\langle\psi|A_{i1}\rangle|B_{j1}\rangle|^2=(\cos{\theta}\cos{\theta}_{ai}\cos{\theta}_{bj}
+\sin{\theta}\sin{\theta_{ai}}\sin{\theta_{bj}})^2.
  \end{array}
\right.
\end{eqnarray}
Take
\begin{eqnarray}
\left\{
  \begin{array}{ll}
    \theta=0.9126956520107119, &  \\
\theta_{a1}=0.7221659141461775,\;\;
\theta_{a2}=1.5035265418613746, &  \\
    \theta_{b1}=1.111854544600825,\;\;
\theta_{b2}=0.30565501091709074, &
  \end{array}
\right.
\end{eqnarray}
then
\begin{eqnarray}\label{err}
\left\{
  \begin{array}{ll}
    \mathcal{H}_1-1=-3.33067\times 10^{-16}, &  \\
    \mathcal{H}_2-1=1.66533\times 10^{-16}, &  \\
     \mathcal{H}_3-1=1.11022\times 10^{-16}, &\\
     P_8=P(A_1=0,B_1=0)\approx 0.391179, &
  \end{array}
\right.
\end{eqnarray}
holds. Thus,  (\ref{eqnp03}) is a general Hardy-type paradox. The numerical calculation error for Hardy's constraints is about $10^{-16}$ (see (\ref{err})).
Hence, in quantum theory, there exist local observables $\{A_1,A_2\}$ and $\{B_1,B_2\}$ and two-qubit state $|\psi\rangle$ such that the three Hardy's constraint in (\ref{eqnp03}) are satisfied, but $P(A_1=0,B_1=0)\approx 0.39>0.$ This completes the proof. $\Box$

{\it Remark 1.}---For the general Hardy-type paradox (\ref{eqnp03}), for simplicity we may initially select some measurement angles as
\begin{eqnarray}
     \theta_{a1}=\frac{\pi}{4},\;\; \theta_{a2}=\frac{\pi}{2},\;\;
   \theta_{b2}=\frac{\pi}{2}-\theta_{b1},
\end{eqnarray}
then we have
\begin{eqnarray}
\left\{
  \begin{array}{ll}
    \mathcal{H}_1-1=\frac{1}{2}\cos^2{(\theta-\theta_{b_1})}+\sin^2{\theta}\sin^2{\theta_{b_1}}-1,&\\
    \mathcal{H}_2-1=\frac{1}{2}\cos^2{(\theta-\theta_{b_1})}+\sin^2{\theta}\sin^2{\theta_{b_1}}-1, &  \\
     \mathcal{H}_3-1=\frac{1}{2}\sin^2{(\theta+\theta_{b_1})}+2\cos^2{\theta}\sin^2{\theta_{b_1}}-1, &\\
     P_8=\frac{1}{2}\sin^2{(\theta+\theta_{b_1})}.
  \end{array}
\right.
\end{eqnarray}
In this case $\mathcal{H}_1=\mathcal{H}_2$. From $\mathcal{H}_1-1=0$ and $\mathcal{H}_3-1=0$, we have

\begin{eqnarray}
    \theta_{b1}=\frac{1}{2}\arccos{x}, \;\;\; \theta=\frac{1}{2}\arccos{\frac{1-x}{-3+5x}},
\end{eqnarray}
where $x$ $(\approx -0.667051)$ satisfies the following algebraic relation
\begin{eqnarray}
73 x^4+28 x^3-70 x^2-12 x+17=0.
\end{eqnarray}
Then in this case we have the successful probability as
%\begin{eqnarray}
%     P_8=\frac{1}{2}\sin^2{(\theta+\theta_{b_1})}=\frac{1}{4}\left(1
%    +\cos{\left(\arcsin{x}+\arcsin{\frac{1-x}{-3+5x}}\right)}\right)\approx 0.385807>0,
%\end{eqnarray}

\begin{eqnarray}
     P_8=\frac{1}{2}\sin^2{(\theta+\theta_{b_1})}=\frac{1}{4}\left[1
    -\cos{\left(\arccos{x}+\arccos{\frac{1-x}{-3+5x}}\right)}\right]\approx 0.385807>0,
\end{eqnarray}
which is very close to the maximum value $0.391179$.

\subsection{Perfect Hardy-type paradox based on the three-qubit MABK inequality}

The three-qubit Mermin-Ardehali-Belinskii-Klyshko (MABK) inequality~\cite{Mermin1990,Ardehali,BK} is given by
\begin{eqnarray}\label{eqMABK}
\langle A_1B_2C_2\rangle+\langle A_2B_1C_2\rangle+\langle A_2B_2C_1\rangle-\langle A_1B_1C_1\rangle\overset{{\rm{LHV}}}{\leq}2\overset{{\rm{QM}}}{\leq}4.
\end{eqnarray}
Here the correlation functions can be written in terms of joint-probabilities as
\begin{eqnarray}\label{qijk}
Q_{ijk}=\langle A_i B_j C_k\rangle=\sum_{a,b,c=0}^1 (-1)^{a+b+c}\; P(A_i=a,B_j=b,C_k=c)=\sum_{r=0}^1 (-1)^r \;P(A_i+B_j+C_k=r)
 \end{eqnarray}
with
\begin{eqnarray}\label{qijk-1}
P(A_i+B_j+C_k=0)&=&P(A_i=0,B_j=0,C_k=0)+P(A_i=1,B_j=1,C_k=0)+\nonumber\\
&&P(A_i=1,B_j=0,C_k=1)+P(A_i=0,B_j=1,C_k=1),\nonumber\\
P(A_i+B_j+C_k=1)&=&P(A_i=1,B_j=0,C_k=0)+P(A_i=0,B_j=1,C_k=0)+\nonumber\\
&&P(A_i=0,B_j=0,C_k=1)+P(A_i=1,B_j=1,C_k=1).
 \end{eqnarray}
Due to
\begin{eqnarray}\label{qijk-2}
P(A_i+B_j+C_k=0)+P(A_i+B_j+C_k=1)=1,
 \end{eqnarray}
we have
\begin{eqnarray}\label{qijk-2a}
\langle A_i B_j C_k\rangle&=&2 P(A_i+B_j+C_k=0)-1\nonumber\\
&=&1-2P(A_i+B_j+C_k=1).
 \end{eqnarray}
Therefore, the three-qubit MABK inequality can be rewritten in terms of joint probabilities as follows
\begin{eqnarray}\label{eqMABK2}
\begin{split}
I_{{\rm{MABK}}}&=P(A_1+B_2+C_2=0)+P(A_2+B_1+C_2=0)+P(A_2+B_2+C_1=0)\\
&~~~~+P(A_1+B_1+C_1=1)\overset{{\rm{LHV}}}{\leq}3\overset{{\rm{QM}}}{\leq}4,
\end{split}
\end{eqnarray}
which is the inequality (8) in the main text.

Based on the three-qubit MABK inequality with the form (\ref{eqMABK2}), we can construct a perfect Hardy-type paradox with success probability equals to 1:
\begin{eqnarray}\label{100paradox}
\left\{
  \begin{array}{ll}
    P(A_1+B_2+C_2=0)=1, &  \\
    P(A_2+B_1+C_2=0)=1, &  \\
     P(A_2+B_2+C_1=0)=1, &\\
     P(A_1+B_1+C_1=1)\overset{{\rm{QM}}}{=}1>0.
  \end{array}
\right.
\end{eqnarray}

\emph{Proof. }In local theory, without loss of generality, we can suppose that the output of the observable $A_1$ is $0$. Then the first constraint in (\ref{100paradox}) implies  $B_2+C_2=0$. For any output of $A_2$, we have $A_2+A_2=0$. Thus, the second and the third constraints in (\ref{100paradox}) lead to $B_1+C_2+B_2+C_1=0$, which yields $B_1+C_1=0$. Therefore, $A_1+B_1+C_1=0$ holds, i.e., $P(A_1+B_1+C_1=1)=0$.

In quantum mechanics, we select the 3-qubit GHZ state in the following form:
\begin{eqnarray}\label{mabk-2}
|\Psi\rangle_{\rm GHZ}=\frac{1}{\sqrt{2}}(|000\rangle-|111\rangle),
 \end{eqnarray}
 and the observables are chosen as
 \begin{eqnarray}\label{ob}
A_1=B_1=C_1=\sigma_{x},\;\;\;A_2=B_2=C_2=\sigma_{y},
 \end{eqnarray}
where $\sigma_x$ and $\sigma_y$ are Pauli matrices. Then we have
\begin{eqnarray}\label{GHZ-four}
\langle A_1 B_2 C_2\rangle&=&\langle\Psi|\sigma_x^A\otimes\sigma_y^B\otimes\sigma_y^C|\Psi\rangle=+1,\nonumber\\
\langle A_2 B_1 C_2\rangle&=&\langle\Psi|\sigma_y^A\otimes\sigma_x^B\otimes\sigma_y^C|\Psi\rangle=+1,\nonumber\\
\langle A_2 B_2 C_1\rangle&=&\langle\Psi|\sigma_y^A\otimes\sigma_y^B\otimes\sigma_x^C|\Psi\rangle=+1,\nonumber\\
\langle A_1 B_1 C_1\rangle&=&\langle\Psi|\sigma_x^A\otimes\sigma_x^B\otimes\sigma_x^C|\Psi\rangle=-1.
\end{eqnarray}
From Eq. (\ref{qijk-2a}), one finds that
the first three Hardy's constraints in (\ref{100paradox}) hold and $P(A_1+B_1+C_1=1)=1$, which completes the proof. $\Box$

{\it Remark 2.}---Due to the relation (\ref{qijk-2}), the three-qubit MABK inequality (\ref{eqMABK2}) can be recast to
\begin{eqnarray}\label{eqMABK3}
I_{{\rm{MABK}}}&=&P(A_1+B_1+C_1=1)\nonumber\\
&&-[P(A_1+B_2+C_2=1)+P(A_2+B_1+C_2=1)+P(A_2+B_2+C_1=1)]\overset{{\rm{LHV}}}{\leq} 0.
\end{eqnarray}
The corresponding general Hardy-type paradox is given by
\begin{eqnarray}\label{100paradox-1}
\left\{
  \begin{array}{ll}
    P(A_1+B_2+C_2=1)=0, &  \\
    P(A_2+B_1+C_2=1)=0, &  \\
     P(A_2+B_2+C_1=1)=0, &\\
     P(A_1+B_1+C_1=1)\overset{{\rm{QM}}}{=}1>0,
  \end{array}
\right.
\end{eqnarray}
which is equivalent to the paradox as shown in (\ref{100paradox}).

\emph{Remark 3.}---The three-qubit GHZ paradox ``$+1=-1$'' is given by (quantum mechanically)~\cite{GHZ89}
\begin{eqnarray}\label{3qubitGHZ}
\sigma_x^A\otimes\sigma_y^B\otimes\sigma_y^C|\Psi\rangle&=&+|\Psi\rangle,\label{GHZ-one}\nonumber\\
\sigma_y^A\otimes\sigma_x^B\otimes\sigma_y^C|\Psi\rangle&=&+|\Psi\rangle,\label{GHZ-two}\nonumber\\
\sigma_y^A\otimes\sigma_y^B\otimes\sigma_x^C|\Psi\rangle&=&+|\Psi\rangle,\label{GHZ-three}\nonumber\\
\sigma_x^A\otimes\sigma_x^B\otimes\sigma_x^C|\Psi\rangle&=&-|\Psi\rangle.\label{GHZ-foura}
\end{eqnarray}
and (classically)
\begin{eqnarray}\label{GHZ-fourP}
v_x^A v_y^B v_y^C&=&+1,\nonumber\\
v_y^A v_x^B v_y^C&=&+1,\nonumber\\
v_y^A v_y^B v_x^C&=&+1.\nonumber\\
v_x^A v_x^B v_x^C&=&-1.
\end{eqnarray}
When one multiplies the four equations in (\ref{GHZ-fourP}), the left-hand side gives $(v_x^A)^2 (v_y^A)^2 (v_x^B)^2 (v_y^B)^2 (v_x^C)^2 (v_y^C)^2=+1$, but the right-hand side gives $-1$, thus leads to the full contradiction of ``$+1=-1$''.

By comparing Eq. (\ref{GHZ-four}) and Eq. (\ref{3qubitGHZ}), one may observe that GHZ paradox (with the full contradiction ``$+1=-1$'') can be viewed as a perfect Hardy-type paradox (with the successful probability being $100\%$). Thus the two apparently different AVN proofs (i.e., GHZ paradox and Hardy's paradox) are unified into a single one (i.e., the general Hardy-type paradox).

\subsection{ General Hardy-type paradox based on the AS inequality}

The Abner Shimony (AS) inequalities are a family of tight two-qubit $n$-setting ($n$ is even) Bell inequalities~\cite{Gisin}\cite{Avis}.  For even number $n$, the AS inequality in terms of correlation functions can be written as
\begin{eqnarray}\label{IN}
	AS_n&=&\sum_{i,j}^{n} \mathcal{M}^n_{ij} \; \langle A_i B_j \rangle\overset{{\rm LHV}}{\leq}\frac{n}{2}\left(\frac{n}{2}+1\right)\overset{{\rm{QM}}}{\leq}\frac{(n+1)\sqrt{n(n+2)}}{3},
\end{eqnarray}
where
$\mathcal{M}^n_{ij}$ is the $i$-th row and $j$-th  column element of the following matrix $\mathcal{M}^n$
\begin{align}\label{eqcij}
		\mathcal{M}^n=\left(
		\begin{array}{c||ccccccccccc}
			& A_1 &A_2 &A_3 & \cdots &A_{\frac{n}{2}} &A_{\frac{n}{2}+1} &A_{\frac{n}{2}+2} & \cdots &A_{n-2} &A_{n-1} & A_{n} \\
			\hline
             \hline
			B_1     & 1   & 1   & 1   & \cdots & 1   & 1       & 1       & \cdots & 1        & 1        & 1 \\
			B_2     & 1   & 1   & 1   & \cdots & 1   & 1       & 1       & \cdots & 1        & 1        & -1 \\
			B_3     & 1   & 1   & 1   & \cdots & 1   & 1       & 1       & \cdots & 1        & -2       & 0 \\
			\vdots  & \vdots& \vdots &\vdots& \cdots & \vdots& \vdots &\vdots& \vdots  & \iddots        &\iddots       & \vdots \\
			B_{\frac{n}{2}}     & 1   & 1   & 1   & \cdots & 1   & 1       & -(\frac{n}{2}-1)  & \cdots & 0        & 0        &0 \\
			B_{\frac{n}{2}+1} & 1   & 1   & 1   & \cdots & 1   & -\frac{n}{2}      & 0       & \cdots & 0        & 0        &0 \\
			B_{\frac{n}{2}+2} & 1   & 1   & 1   & \cdots & -(\frac{n}{2}-1)& 0       & 0  & \cdots & 0        & 0        &0 \\
			\vdots  & \vdots& \vdots &\vdots& \cdots & \vdots& \vdots &\vdots& \vdots     & \vdots &\vdots& \vdots \\
			B_{n-2}& 1   & 1   & 1    & \iddots & 0        & 0        &0      & \cdots & 0 & 0 & 0 \\
			B_{n-1}& 1   & 1   & -2    & \iddots & 0        & 0        &0       & \cdots      & 0 & 0 & 0 \\
			B_{n}  & 1   & -1   & 0    & \cdots & 0        & 0        &0       &\cdots      & 0 & 0 & 0\\
		\end{array}
		\right).
	\end{align}	
 Due to the relation (\ref{qij-1}), the AS inequality can also be rewritten in terms of joint-probabilities as (for convenient, here we use the notation $\mathcal{I}_{nn22}$ to represent the Bell function $AS_n$):
\begin{align}\label{IN2-0}
\begin{split}
\mathcal{I}_{nn22}&=\sum_{i=1}^{n}\sum_{j=1}^{n-i+1} P(A_i=B_j)+\sum_{i=2}^{\frac{n}{2}}(i-1)[P(A_i\neq B_{n-i+2})+P(A_{n+2-i}\neq B_{i})]+\frac{n}{2} P(A_{\frac{n}{2}+1}\neq B_{\frac{n}{2}+1})\\
&\overset{{\rm LHV}}{\leq}\frac{n^2+n}{2}\\
&\overset{{\rm{QM}}}{\leq}\frac{\frac{(n+1)\sqrt{n(n+2)}}{3}+\frac{3n^2+2n}{4}}{2}.
\end{split}
\end{align}

\emph{Remark 4.---}For $n=2$, the AS inequality $\mathcal{I}_{2222}\leq 3$ is nothing but the CHSH inequality $\mathcal{I}_{{\rm{CHSH}}}\leq 3$  given in (\ref{eqsp}). Namely, the inequality $\mathcal{I}_{nn22}\leq \frac{n^2+n}{2}$ is a generalization of CHSH inequality from two-setting to $n$-setting. In the previous section and the main text, we have presented the general Hardy-type paradox based on the inequality (\ref{eqsp}). The corresponding successful probability is about $0.391179$, which is about four times of $9\%$ (the successful probability for the original Hardy's paradox~\cite{hardy93}).

\emph{Remark 5.---}To show more examples for the general Hardy-type paradox based on Bell inequality, here we study the AS inequality for $n=4$. The inequality reads
% \begin{widetext}
\begin{align}\label{IN2}
\begin{split}
	\mathcal{I}_{4422}&=\sum_{i=1}^{26}f_iP_i\\
&=P(A_1=0,B_2=0)+P(A_2=1,B_3=1)+P(A_1=1,B_3=1)+2 P(A_{3}=1, B_{3}=0)\\
&~~~~+P(A_2=1,B_2=1)+P(A_3=0,B_1=0)+P(A_2=1, B_{4}=0)+P(A_4=0,B_1=0)\\
&~~~~+P(A_3=1,B_2=1)+2 P(A_{3}=0,B_{3}=1)\\
&~~~~+P(A_1=0,B_4=0)+P(A_1=1,B_4=1)+P(A_2=1,B_1=1)\\
&~~~~+P(A_2=0,B_1=0)+P(A_2=0,B_3=0)+P(A_3=1,B_1=1)\\
&~~~~+P(A_2=0,B_2=0)+P(A_2=0, B_{4}=1)+P(A_4=1,B_1=1)\\
&~~~~+P(A_1=1,B_1=1)+P(A_1=1,B_2=1)+P(A_1=0,B_3=0)+P(A_3=0,B_2=0)\\
&~~~~+P(A_{4}=0,B_{2}=1)+P(A_{4}=1,B_{2}=0)+P(A_1=0,B_1=0)\\
&\overset{{\rm LHV}}\leq 10\\
&\overset{{\rm{QM}}}\leq 7+\frac{5\sqrt{6}}{3}.
\end{split}
\end{align}
The inequality contains 26 probabilities, and the corresponding Hardy-type paradox can be given by
\begin{eqnarray}\label{eqI4422}
\left\{
  \begin{array}{ll}
    \mathcal{H}_1=P_1+P_2=1, &  \\
    \mathcal{H}_2=P_3+2P_4=1, & \\
\mathcal{H}_3=P_5+P_6=1, & \\
\mathcal{H}_4=P_7+P_8=1, & \\
\mathcal{H}_5=P_9+2P_{10}=1, & \\
\mathcal{H}_6=P_{11}+P_{12}+P_{13}=1, & \\
\mathcal{H}_7=P_{14}+P_{15}+P_{16}=1, & \\
\mathcal{H}_8=P_{17}+P_{18}+P_{19}=1, & \\
\mathcal{H}_9=P_{20}+P_{21}+P_{22}+P_{23}+P_{24}+P_{25}=2,& \\
P_{26}>0. &
  \end{array}
\right.
\end{eqnarray}

{\it Proof.} In local theory, similarly, if the first $9$ Hardy's constraints in (\ref{eqI4422}) hold,
then we can obtain
\begin{eqnarray}
A_1=B_1=A_2=B_2=A_3=B_3=A_4=1,B_4=0,
\end{eqnarray}
 which yields  $P_{26}=P(A_1=0,B_1=0)=0$.

In quantum mechanics, we take the local observables $\{A_1,A_2,A_3,A_4\}$, $\{B_1,B_2,B_3,B_4\}$ and the two-qubit state $|\psi\rangle$ to be the forms as in Eq. (\ref{eqmAB}). After substituting the following measurement angles
\begin{eqnarray}
\left\{
  \begin{array}{ll}
    \theta=0.595078,&  \\
\theta_{a1}=1.429925, \theta_{a2}=0.710013,
\theta_{a3}=2.050002, \theta_{a4}=2.113850, &  \\
    \theta_{b1}=1.641220, \theta_{b2}=0.958109,
 \theta_{b3}=0.807975, \theta_{b4}=-1.379539, &
  \end{array}
\right.
\end{eqnarray}
into Eq. (\ref{eqI4422}), then the following relation
\begin{eqnarray}\label{eqI44222}
\left\{
  \begin{array}{ll}
    \mathcal{H}_1-1\approx-5.45939\times 10^{-7}, &  \\
    \mathcal{H}_2-1\approx8.07675\times 10^{-7}, &  \\
\mathcal{H}_3-1\approx-3.93129\times 10^{-7}, &\\
\mathcal{H}_4-1\approx-2.90784\times 10^{-7}, &\\
\mathcal{H}_5-1\approx-3.62443\times 10^{-7}, &\\
\mathcal{H}_6-1\approx4.62613\times 10^{-7}, &\\
\mathcal{H}_7-1\approx3.65377\times 10^{-7}, &\\
\mathcal{H}_8-1\approx5.81279\times 10^{-7}, &\\
\mathcal{H}_9-2\approx10^{-7}, &\\
P_{26}=P(A_1=0,B_1=0)\approx 0.659882>0,
  \end{array}
\right.
\end{eqnarray}
holds. The proof is completed. $\Box$

 In this case, the maximal successful probability is about $0.659$, which is about 7 times of $9\%$ (the successful probability for the original Hardy's paradox~\cite{hardy93}).

\subsection{The ladder proof of nonlocality without inequalities}

For convenience, here we briefly review the ladder proof of nonlocality without inequalities in Ref. \cite{Hardy1997}, which is a kind of extension of Hardy's paradox from two-setting to $k$-setting. For $k=2$, the paradox reduces to the original Hardy's paradox in~\cite{hardy93}.

According to quantum mechanics, there always exist two-qubit entangled states and
local measurements satisfying, simultaneously,
 \begin{equation}\label{eqH1}
\left\{
  \begin{array}{ll}
    P(A_k=1,B_k=1)>0, \\
    P(A_i=1,B_{i-1}=0)=0, \\
    P(A_{i-1}=0,B_{i}=1)=0, \\
    P(A_1=1,B_1=1)=0,
  \end{array}
\right.
 \end{equation}
 for any $i=2,\cdots,k$. However, if  event  $A_k=1,B_k=1$  happens, then, in any local theory, the event $A_1=1,B_1=1$ must happen by the constraint conditions $P(A_i=1,B_{i-1}=0)=0$ and $P(A_{i-1}=0,B_{i}=1)=0$  for any $i=2,\cdots,k$, which contracts with the last constraint condition in (\ref{eqH1}).
Because of the invariance of  $\max\{P(A_k=1,B_k=1)\}$ under locally unitary operations, it is sufficient to  consider
the pure state  $|\psi\rangle=\alpha|00\rangle-\beta|11\rangle,$
with $\alpha,\beta\geq0$ and $\alpha^2+\beta^2=1$, for which the optimal observables, i.e., they satisfy the constraint conditions in (\ref{eqH1}) and lead to the maximum probability of the nonlocal event, are
 \begin{eqnarray}
 &&|A_{i1}\rangle=|B_{i1}\rangle=\frac{1}{\alpha^{2i-1}+\beta^{2i-1}}\left(
                              \begin{array}{c}
                                (-1)^i\beta^{i-\frac{1}{2}}\\
                                \alpha^{i-\frac{1}{2}} \\
                              \end{array}
                            \right),\nonumber\\
                            &&|A_{i0}\rangle\bot |A_{i1}\rangle,\;\;\;|B_{i0}\rangle\bot |B_{i1}\rangle,\;\;\;(i=1,2,\cdots,k),
                            \end{eqnarray}
and in this case,
 \begin{eqnarray}
 &&P(A_k=1,B_k=1)=|\langle\psi|A_{i1}\rangle|B_{i1}\rangle|^2=\left(\frac{\alpha \beta^{2k-1}-\beta \alpha^{2k-1}}{\beta^{2k-1}+\alpha^{2k-1}}\right)^2
  \end{eqnarray}
holds. Therefore, we obtain the successful probability as
 \begin{eqnarray}
P_{SC}(k)=\max\{P(A_k=1,B_k=1)\}=\left\{\left(\frac{\alpha \beta^{2k-1}-\beta \alpha^{2k-1}}{\beta^{2k-1}+\alpha^{2k-1}}\right)^2:\alpha,\beta\geq0,\alpha^2+\beta^2=1\right\}.
 \end{eqnarray}

\begin{figure}[hbtp]%% t: top, b: bottom, p: in a page containing only floats, and h: here.
  \centering{\includegraphics[width=75mm]{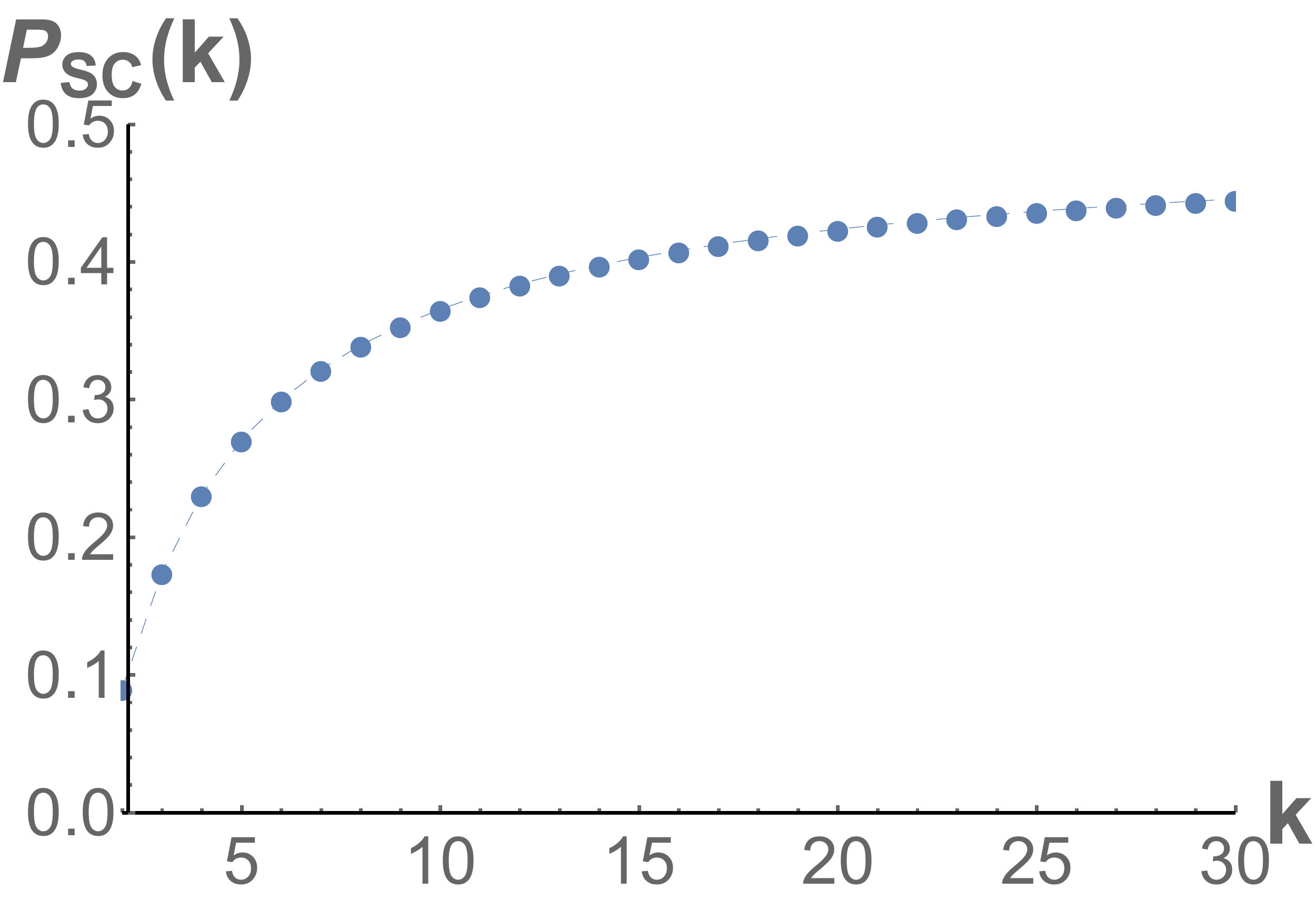}}
  \caption{\label{fig1}The relationship between $P_{SC}(k)$ and $k$.}
\end{figure}

{\it Remark 6.}---The result of Ref. \cite{Hardy1997} shows that $P_{SC}(k)$ grows with increasing $k$, and $P_{SC}(k)\rightarrow 0.5$ as $k\rightarrow \infty$.
In Table \ref{table1}, we list $P_{SC}(k)$ for $k=2,3,4,5,6$. Furthermore, we plot the relationship between $P_{SC}(k)$ and $k$ in Fig. \ref{fig1} with $2\leq k\leq 30$.

\begin{table}[hbtp]
\caption{\label{table1}  $P_{SC}(k)$ versus $k$ for $k=2,3,4,5,6$ (The numerical result given in Ref. \cite{Hardy1997}).}
\begin{ruledtabular}
\begin{tabular}{cccccc}
 &  $k=2$ & $k=3$ & $k=4$& $k=5$ & $k=6$\\
  \hline
  $P_{SC}(k)$ &  $0.09017$ & $0.17455$ & $0.23126$& $0.27088$ & $0.29995$\\
\end{tabular}
\end{ruledtabular}
\end{table}

In a real experiment, inequalities are necessary to show
that the errors do not wash out the logical contradiction
that local realism faces.  Based on the ladder Hardy's paradox as shown in (\ref{eqH1}), in addition Ref. \cite{Hardy1997} established a family of Bell inequalities as
 \begin{eqnarray}\label{eqLadder1}
\mathcal{I}_{k}^{\rm ladder} &=&P(A_k=1,B_k=1)-\sum_{i=2}^kP(A_i=1,B_{i-1}=0)-\sum_{i=2}^kP(A_{i-1}=0,B_{i}=1)\nonumber\\
&&-P(A_1=1,B_1=1)\leq 0,\;\;\; \forall k\geq 2.
\end{eqnarray}
Inversely, based on Bell inequalities in (\ref{eqLadder1}), one can derive the Hardy-type paradox as in (\ref{eqH1}).

%In fact, one can generalize them to the following  ones:
% \begin{equation}\label{eqLadder2}
%P(A_k=1,B_k=1)-\sum_{i=2}^k x_iP(A_i=1,B_{i-1}=0)-\sum_{i=2}^k y_i P(A_{i-1}=0,B_{i}=1)-zP(A_1=1,B_1=1)\leq 0,\forall k\geq 2.
%\end{equation}
%with $x_i,y_i,z$ being positive integers.

\subsection{General Hardy-type paradoxes based on $(2k+1)$-cycle Bell inequalities}

\begin{figure}[hbtp]%% t: top, b: bottom, p: in a page containing only floats, and h: here.
  \centering{\includegraphics[width=95mm]{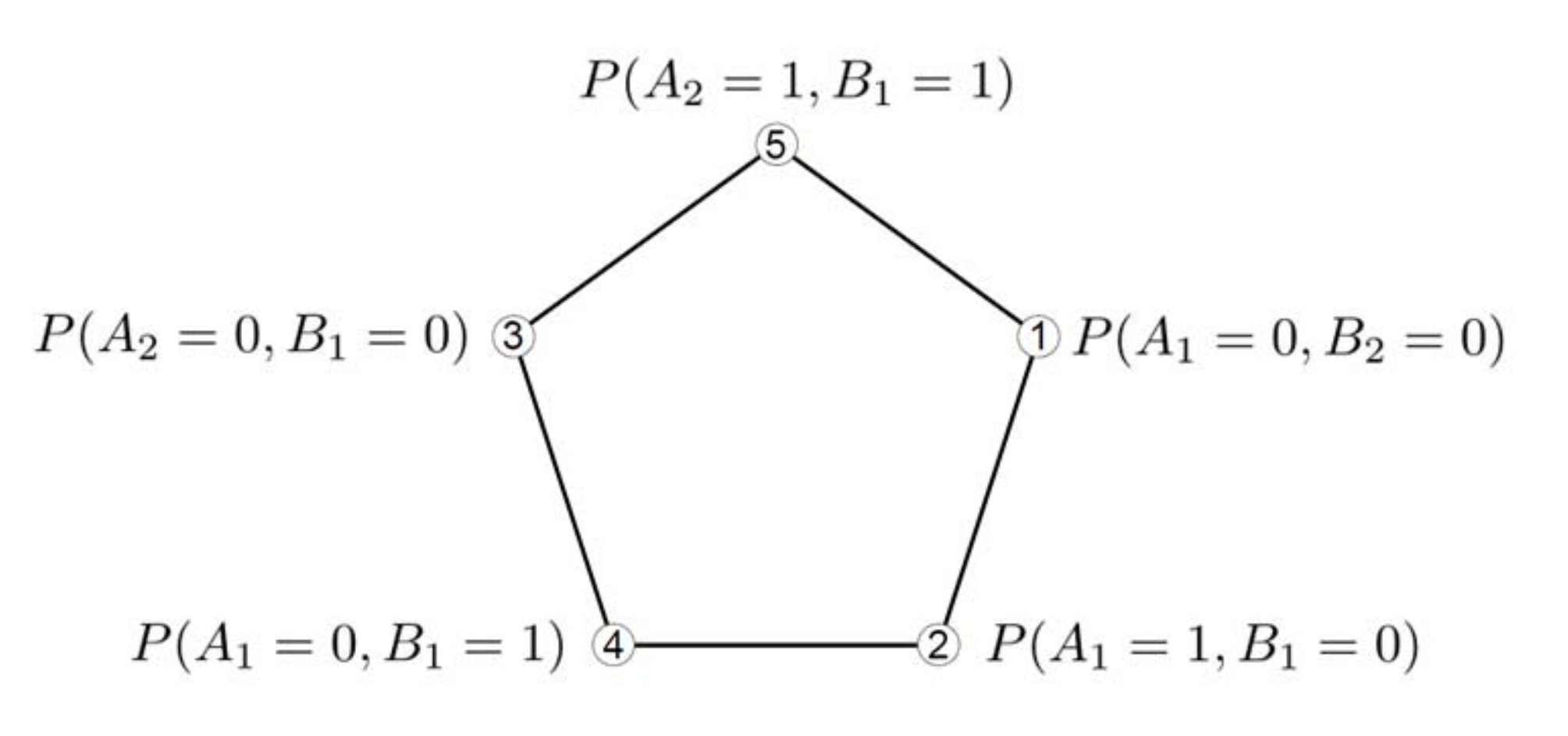}}
  \caption{\label{fig2}The $5$-cycle exclusivity graph, where two vertices are connected if and only if the
events represented by them are exclusive. For example, vertex 1 and vertex 2 are connected because the event $(A_1=0,B_2=0)$ and the event $(A_1=1,B_1=0)$ are exclusive, etc.}
\end{figure}

 The simplest exclusivity graph \cite{Cabello} is the $5$-cycle exclusivity graph (see Fig. \ref{fig2}).

One kind of  Bell inequalities that generated from the $5$-cycle exclusivity graph (corresponds to $k=2$) is given by
\begin{eqnarray}\label{eq(8)}
%\begin{split}
&&\mathcal{I}_{k=2}^{\rm cycle}=\sum_{i=1}^5P_i\nonumber\\
&&=P(A_1=0,B_2=0)+P(A_1=1,B_1=0)+P(A_2=0,B_1=0)+P(A_1=0,B_1=1)+P(A_2=1,B_2=1)\nonumber\\
&&\overset{{\rm{LHV}}}\leq 2.
%\end{split}
\end{eqnarray}
For the $5$-cycle Bell inequality, we can construct a general Hardy-type paradox as follows:
\begin{equation}\label{eqfnp}
\left\{
  \begin{array}{ll}
    \mathcal{H}_1=P(A_1=0,B_2=0)+P(A_1=1,B_1=0)=P_1+P_2=1, &  \\
    \mathcal{H}_2=P(A_2=0,B_1=0)+P(A_1=0,B_1=1)=P_3+P_4=1, &   \\
P_5=P(A_2=1,B_2=1)>0.&
  \end{array}
\right.
 \end{equation}

{\it Proof.} It is easy to have
\begin{eqnarray}\label{EE1}
&&    P(A_1=0,B_2=0)+P(A_1=0,B_2=1)+P(A_1=1,B_1=0)+P(A_1=1,B_1=1)\nonumber\\
&&    =P(A_1=0)+P(A_1=1)=1,
 \end{eqnarray}
\begin{eqnarray}\label{EE2}
  &&  P(A_2=0,B_1=0)+P(A_2=1,B_1=0)+P(A_1=0,B_1=1)+P(A_1=1,B_1=1)\nonumber\\
  &&=P(B_1=0)+P(B_1=1)=1.
\end{eqnarray}

By calculating Eq. (\ref{EE1})$+$ Eq. (\ref{EE2})$-\mathcal{H}_1-\mathcal{H}_2=0$, we immediately finds that  the constraint conditions in (\ref{eqfnp}) hold if and only if the following conditions
\begin{eqnarray}\label{oHc}
P(A_2=1,B_1=0)=0,\;\; P(A_1=0,B_2=1)=0,\;\; P(A_1=1,B_1=1)=0,
\end{eqnarray}
hold. Note that
the conditions in (\ref{oHc}) are just the constraint conditions of the paradox (\ref{eqH1}) (for $k=2$) and the nonlocal event $P(A_2=1,B_2=1)$ in paradox (\ref{eqfnp}) is just the one in the paradox (\ref{eqH1}). Thus, (\ref{eqfnp}) is a general Hardy-type paradox, which is equivalent to the paradox (\ref{eqH1}). $\Box$

{\it Remark 7.}---According to the above proof, the general Hardy-type paradox (\ref{eqfnp}) based on $5$-cycle Bell inequality
is   equivalent to the ladder Hardy's paradox given in (\ref{eqH1}) with $k=2$.

The above discussion can be generalized to the general Hardy-type paradoxes based on $(2k+1)$-cycle Bell inequalities, which can be generated from the $(2k+1)$-cycle exclusivity graph. In the following, we present details for the construction.

 \begin{itemize}
 \item $k$ is an odd number: For any joint assignment of all measurements $\{A_i,B_j\}$, one Bell inequality according to the $(2k+1)$-cycle exclusivity graph is described by
\begin{eqnarray}
\mathcal{I}_{k}^{\rm cycle}&=&\sum_{i=1}^{2k+1}P_i\nonumber\\
&&=P(A_{k-1}=0,B_k=0)+P(A_{k-1}=1,B_{k-2}=1)+P(A_{k}=0,B_{k-1}=0)+P(A_{k-2}=1,B_{k-1}=1)\nonumber\\
&&+P(A_{k-3}=0,B_{k-2}=0)+P(A_{k-3}=1,B_{k-4}=1)+P(A_{k-2}=0,B_{k-3}=0)+P(A_{k-4}=1,B_{k-3}=1)\nonumber\\
&&+P(A_{k-5}=0,B_{k-4}=0)+P(A_{k-5}=1,B_{k-6}=1)+P(A_{k-4}=0,B_{k-5}=0)+P(A_{k-6}=1,B_{k-5}=1)\nonumber\\
&&+\cdots\nonumber\\
&&+P(A_{2}=0,B_{3}=0)+P(A_{2}=1,B_{1}=1)+P(A_{3}=0,B_{2}=0)+P(A_{1}=1,B_{2}=1)\nonumber\\
&&+P(A_{2}=0,B_{1}=0)+P(A_{1}=0,B_{1}=1)+P(A_k=1,B_k=1)\nonumber\\
&&\overset{{\rm LHV}}\leq k.
\end{eqnarray}
The general Hardy-type paradox based on the $(2k+1)$-cycle Bell inequality is given by
\begin{eqnarray}\label{eqS4}
\left\{
  \begin{array}{ll}
    \mathcal{H}_1=P(A_{k-1}=0,B_k=0)+P(A_{k-1}=1,B_{k-2}=1)=P_1+P_2=1, \\
    \mathcal{H}_2=P(A_{k}=0,B_{k-1}=0)+P(A_{k-2}=1,B_{k-1}=1)=P_3+P_4=1, \\
    \mathcal{H}_3=P(A_{k-3}=0,B_{k-2}=0)+P(A_{k-3}=1,B_{k-4}=1)=P_5+P_6=1, \\
    \ \ \ \ \ \ \ \ \ \ \ \ \ \ \ \ \ \ \ \ \ \ \ \ \ \ \ \ \ \ \ \ \ \ \ \ \ \ \ \ \ \ \vdots \\
\mathcal{H}_{k-2}=P(A_{2}=0,B_{3}=0)+P(A_{2}=1,B_{1}=1)=P_{2k-5}+P_{2k-4}=1, \\
    \mathcal{H}_{k-1}=P(A_{3}=0,B_{2}=0)+P(A_{1}=1,B_{2}=1)=P_{2k-3}+P_{2k-2}=1, \\
\mathcal{H}_{k}=P(A_{2}=0,B_{1}=0)+P(A_{1}=0,B_{1}=1)=P_{2k-1}+P_{2k}=1, \\
    P_{2k+1}=P(A_k=1,B_k=1)>0.
  \end{array}
\right.
\end{eqnarray}

{\it Proof.} By
\begin{eqnarray}
\left\{
  \begin{array}{ll}
    P(A_{k-1}=0)+P(A_{k-1}=1)=1, \\
    P(B_{k-1}=0)+P(B_{k-1}=1)=1, \\
    P(A_{k-3}=0)+P(A_{k-3}=1)=1, \\
P(B_{k-3}=0)+P(B_{k-3}=1)=1, \\
    \ \ \ \ \ \ \ \ \ \ \ \ \ \ \ \vdots \\
P(A_{2}=0)+P(A_{2}=1)=1, \\
 P(B_{2}=0)+P(B_{2}=1)=1, \\
P(B_{1}=0)+P(B_{1}=1)=1,
  \end{array}
\right.
\end{eqnarray}
we obtain that the first $k$ Hardy's constraints in (\ref{eqS4}) are equivalent to the $k$ constraint conditions in the ladder Hardy's paradox as shown in (\ref{eqH1}).
 Moreover,
the nonlocal event $P(A_k=1,B_k=1)$ is also the one in the ladder Hardy's paradox. Therefore, by Ref. \cite{Hardy1997}, (\ref{eqS4})
is a general Hardy-type paradox, which is equivalent to the ladder proof of Hardy's paradox  in (\ref{eqH1}). $\Box$

\item $k$ is an even number: For any joint assignment of all measurements $\{A_i,B_j\}$, it holds that
\begin{eqnarray}
\mathcal{I}_{k}^{\rm cycle}&=&P(A_{k-1}=0,B_k=0)+P(A_{k-1}=1,B_{k-2}=1)+P(A_{k}=0,B_{k-1}=0)+P(A_{k-2}=1,B_{k-1}=1)\nonumber\\
&&+P(A_{k-3}=0,B_{k-2}=0)+P(A_{k-3}=1,B_{k-4}=1)+P(A_{k-2}=0,B_{k-3}=0)+P(A_{k-4}=1,B_{k-3}=1)\nonumber\\
&&+P(A_{k-5}=0,B_{k-4}=0)+P(A_{k-5}=1,B_{k-6}=1)+P(A_{k-4}=0,B_{k-5}=0)+P(A_{k-6}=1,B_{k-5}=1)\nonumber\\
&&+\cdots\nonumber\\
&&+P(A_{1}=0,B_{2}=0)+P(A_{1}=1,B_{1}=0)+P(A_{2}=0,B_{1}=0)+P(A_{1}=0,B_{1}=1)\nonumber\\
&&+P(A_k=1,B_k=1)\nonumber\\
&\overset{LHV}\leq k.
\end{eqnarray}
The corresponding general Hardy-type paradox is given by
\begin{eqnarray}\label{eqS5}
\left\{
  \begin{array}{ll}
    \mathcal{H}_1=P(A_{k-1}=0,B_k=0)+P(A_{k-1}=1,B_{k-2}=1)=P_1+P_2=1, \\
    \mathcal{H}_2=P(A_{k}=0,B_{k-1}=0)+P(A_{k-2}=1,B_{k-1}=1)=P_3+P_4=1, \\
    \mathcal{H}_3=P(A_{k-3}=0,B_{k-2}=0)+P(A_{k-3}=1,B_{k-4}=1)=P_5+P_6=1, \\
    \ \ \ \ \ \ \ \ \ \ \ \ \ \ \ \ \ \ \ \ \ \ \ \ \ \ \ \ \ \ \ \ \ \ \ \ \ \ \ \ \ \ \vdots \\
    \mathcal{H}_{k-1}=P(A_{1}=0,B_{2}=0)+P(A_{1}=1,B_{1}=0)=P_{2k-3}+P_{2k-2}=1, \\
\mathcal{H}_{k}=P(A_{2}=0,B_{1}=0)+P(A_{1}=0,B_{1}=1)=P_{2k-1}+P_{2k}=1, \\
    P_{2k+1}=P(A_k=1,B_k=1)>0.
  \end{array}
\right.
\end{eqnarray}

{\it Proof.} Due to
\begin{eqnarray}
\left\{
  \begin{array}{ll}
    P(A_{k-1}=0)+P(A_{k-1}=1)=1, \\
    P(B_{k-1}=0)+P(B_{k-1}=1)=1, \\
    P(A_{k-3}=0)+P(A_{k-3}=1)=1, \\
P(B_{k-3}=0)+P(B_{k-3}=1)=1, \\
    \ \ \ \ \ \ \ \ \ \ \ \ \ \ \ \vdots \\
P(A_{1}=0)+P(A_{1}=1)=1, \\
 P(B_{1}=0)+P(B_{1}=1)=1,
  \end{array}
\right.
\end{eqnarray}
we have that
the first $k$ constraint conditions in (\ref{eqS5}) are equivalent to the $k$ constraint conditions in the ladder Hardy's paradox as shown in (\ref{eqH1}).
 Moreover, the nonlocal event $P(A_k=1,B_k=1)$ is also the one in the ladder Hardy's paradox. Therefore, by Ref. \cite{Hardy1997}, (\ref{eqS5})
is a general Hardy-type paradox, which is equivalent to the ladder proof of Hardy's paradox  in (\ref{eqH1}). $\Box$
\end{itemize}

{\it Remark 8.}---According to the above proof, the general Hardy-type paradox based on $(2k+1)$-cycle Bell inequality
is   equivalent to the ladder Hardy's paradox given in (\ref{eqH1}).

%{\it Remark 8}---For odd $k$, the proof of Example 5 yields
%\begin{eqnarray*}
%& &I_k\\
%&=&(1-P(A_{k-1}=0,B_k=1)-P(A_{k-1}=1,B_{k-2}=0))+(1-P(A_{k}=1,B_{k-1}=0)-P(A_{k-2}=0,B_{k-1}=1))\\
%& &+(1-P(A_{k-3}=0,B_{k-2}=1)-P(A_{k-3}=1,B_{k-4}=0))+(1-P(A_{k-2}=1,B_{k-3}=0)-P(A_{k-4}=0,B_{k-3}=1))\\
%& &+\cdots\\
%& &+(1-P(A_{2}=0,B_{3}=1)-P(A_{2}=1,B_{1}=0))+(1-P(A_{3}=1,B_{2}=0)-P(A_{1}=0,B_{2}=1))\\
%& &+(1-P(A_{2}=1,B_{1}=0)-P(A_{1}=1,B_{1}=1))+P(A_{k}=1,B_{k}=1)\\
%&=&k-P(A_{k}=1,B_{k-1}=0)-P(A_{k-1}=0,B_k=1)-P(A_{k-1}=1,B_{k-2}=0)-P(A_{k-2}=0,B_{k-1}=1)\\
%& &-P(A_{k-2}=1,B_{k-3}=0)-P(A_{k-3}=0,B_{k-2}=1)-P(A_{k-3}=1,B_{k-4}=0)-P(A_{k-4}=0,B_{k-3}=1)\\
%& &-\cdots\\
%& &-P(A_{3}=1,B_{2}=0)-P(A_{2}=0,B_{3}=1)-P(A_{2}=1,B_{1}=0)-P(A_{1}=0,B_{2}=1)\\
%& &-P(A_{2}=1,B_{1}=0)-P(A_{1}=1,B_{1}=1)+P(A_{k}=1,B_{k}=1)\\
%&=&P(A_{k}=1,B_{k}=1)-\sum_{i=3}^kP(A_{i}=1,B_{i-1}=0)-\sum_{i=2}^kP(A_{i-1}=0,B_{i}=1)\\
%& &-2P(A_{2}=1,B_{1}=0)-P(A_{1}=1,B_{1}=1)+k,
%\end{eqnarray*}
%which leads to
%$I_k\leq k$ if and only if
%$$P(A_{k}=1,B_{k}=1)-
%2P(A_{2}=1,B_{1}=0)-\sum_{i=3}^kP(A_{i}=1,B_{i-1}=0)-\sum_{i=2}^kP(A_{i-1}=0,B_{i}=1)-P(A_{1}=1,B_{1}=1)\leq0,$$
%which is the Bell inequality (\ref{eqLadder2}) with $x_2=2,x_i=1,y_j=1,z=1$ for $i=3,4,\cdots,k;j=2,3,\cdots,k.$ Similarly, for even $k$,  the Bell inequality $I_k\leq k$ is equivalent to some Bell inequality (\ref{eqLadder2}) with suitable positive integers $x_i,y_i,z$.

\section{The experimental part}

\subsection{Experimental measurement of probabilities}

\begin{figure}[t]
	\begin{center}
		\includegraphics[width=80mm]{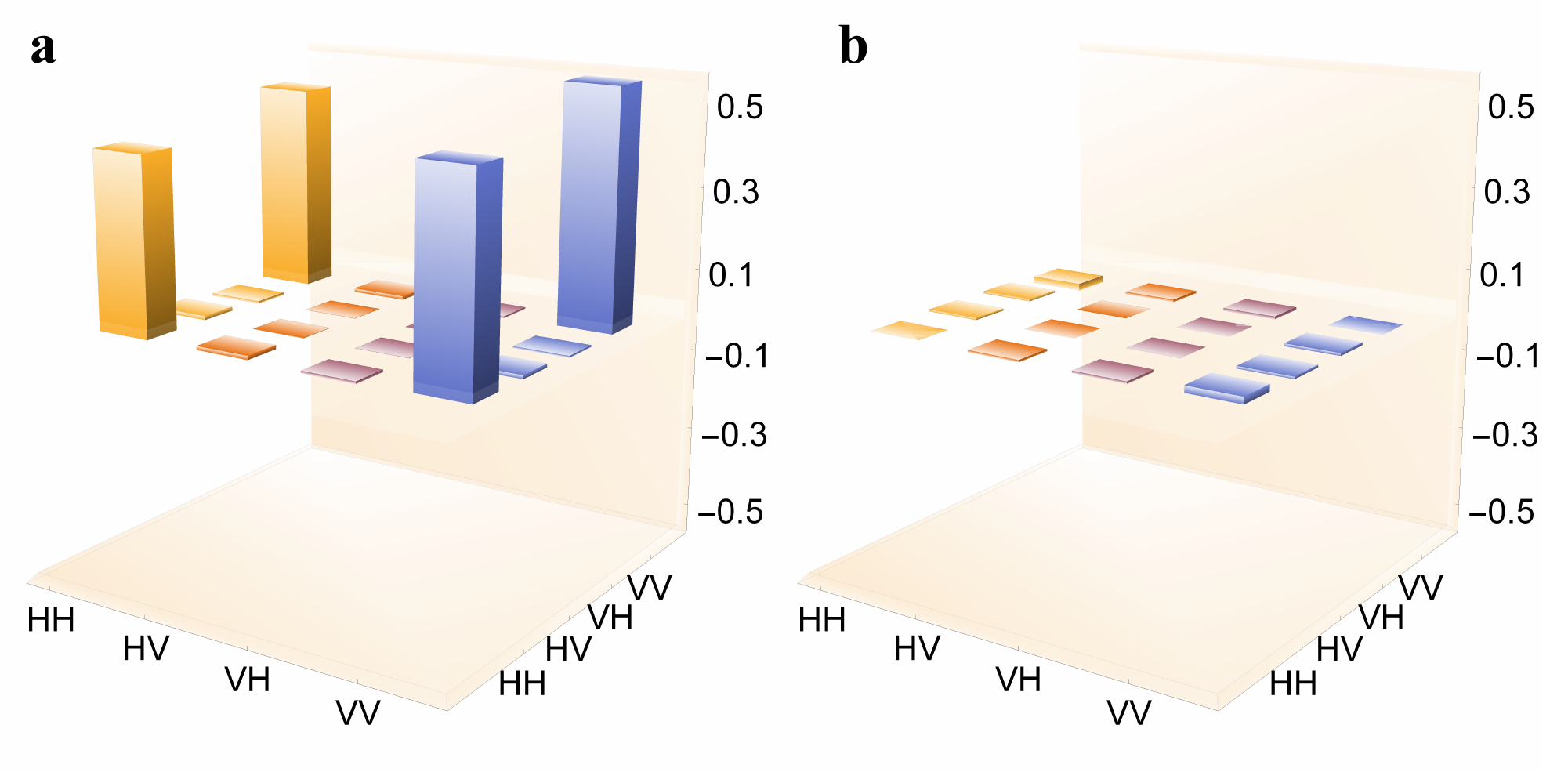}
		\caption{(color online). The real part \textbf{a.} and imaginary part \textbf{b.} of the prepared two-qubit density matrix for verifying the Hardy-type paradox based on the Bell inequality $\mathcal{I}_{\rm CHSH}\leq 3$.}
		\label{tomo_P8}
	\end{center}
\end{figure}

To show the max conflict of the general Hardy-type paradox based on the Bell inequality $\mathcal{I}_{\rm CHSH}\leq 3$, we set $\theta= 0.912$ for the two-qubit state $|\psi\rangle= \cos\theta|HH\rangle+\sin\theta|VV\rangle$. The tomography result is shown in Fig. \ref{tomo_P8}, and the fidelity is over $0.99$. To measure the probabilities, we directly measure coincident counts in 16 different settings, which are shown in Table \ref{Proj8}.
To clearly show the confliction, the projective measurement on Alice side $A_{i1}$ is chosen to be $\cos\theta_{ai}|H\rangle+\sin\theta_{ai}|V\rangle$ with $\theta_{ai}$=0.722 and 1.504 ($i=1, 2$), respectively.The projective measurement on Bob's side $B_{i1}$ is chosen to be $\cos\theta_{bi}|H\rangle+\sin\theta_{bi}|V\rangle$, with $\theta_{bi}$=1.112 and 0.306 ($i=1, 2$).

\begin{table}[htbp]
	\caption{\label{tab:test} Results of Projective Measurements\tnote{1}}
	\begin{ruledtabular}
	\begin{tabular}{cccccccc}		
		Alice, Bob  & Coincidence  & Alice, Bob & Coincidence & Alice, Bob & Coincidence & Alice, Bob & Coincidence\\
		Settings & Counts  & Settings & Counts  & Settings & Counts  & Settings & Counts \\
		\hline
		$HH$ & $N_{1}=18697$ & $A_{11}B_{11}$ & $N_{5}=22897$ & $HB_{11}$ & $N_{9}=4429$ & $A_{11}H$ & $N_{13}=10212$ \\
		$HV$ & $N_{2}=51$ & $A_{21}B_{11}$ & $N_{6}=25642$ & $VB_{11}$ & $N_{10}=23961$ & $A_{11}V$ & $N_{14}=14124$ \\
		$VH$ & $N_{3}=42$ & $ A_{11}B_{21}$ & $N_{7}=18013$ & $HB_{21}$ & $N_{11}=18372$ & $A_{21}H$ & $N_{15}=47$\\
		$VV$ & $N_{4}=32330$ & $A_{21}B_{21}$ & $N_{8}=3131$ & $VB_{21}$ & $N_{12}=2521$ & $A_{21}V$ & $N_{16}=30718$ \\
		%$A_{11}B_{11}$ & $N_{5})$ & $A_{11}H$ & $N_{13})$ \\
		%$A_{21}B_{11}$ & $N_{6})$ & $A_{11}V$ & $N_{14})$ \\
		%$A_{11}B_{21}$ & $N_{7})$ & $A_{21}H$ & $N_{15})$ \\
		%$A_{21}B_{21}$ & $N_{8})$ & $A_{21}V$ & $N_{16})$ \\		
	\end{tabular}\label{Proj8}
\end{ruledtabular}
	\begin{tablenotes}		
		\footnotesize
		\item[1] $|\psi\rangle= \cos\theta|HH\rangle+\sin\theta|VV\rangle$; $A_{i1}= \cos\theta_{ai}\vert H\rangle+\sin\theta_{ai}\vert V\rangle$;
		$B_{i1}= \cos\theta_{bi}\vert H\rangle+\sin\theta_{bi}\vert V\rangle$;\\
$\theta= 0.912, \;\;\theta_{a1}=0.722,\;\; \theta_{a2}=1.504,\;\; \theta_{b1}=1.112,\;\;\theta_{b2}=0.306$.
	\end{tablenotes}
\end{table}

% The coincidence window is 3ns, and the total number of photons are collected within the integral time 1 second. %is 10484. We split every probability into a sum of multiple probabilities, and seek the probability of each item separately.

 In this section, we give an example on how we calculate the probability $P_{8}$. All the other probabilities can be calculated with the similar method. The probability $P_{8}$ is represented by:
\begin{equation}
P_{8}=\langle\psi|(|A_{10}\rangle\langle A_{10}|)\otimes(|B_{10}\rangle\langle B_{10}|)|\psi\rangle=\langle\psi|(I_{2}-|A_{11}\rangle\langle A_{11}|)\otimes(I_{2}-|B_{11}\rangle\langle B_{11}|)|\psi\rangle,
\end{equation}
which can be decomposed into $\langle\psi|I_{2}\otimes I_{2}|\psi\rangle$,
$-\langle\psi|A_{11}\rangle\langle A_{11}|\otimes I_{2}|\psi\rangle$,
$-\langle\psi| I_{2}\otimes|B_{11}\rangle\langle B_{11}|\psi\rangle$
and $\langle\psi|A_{11}\rangle\langle A_{11}|\otimes|B_{11}\rangle\langle B_{11}|\psi\rangle$.
Moreover, $\langle\psi|I_{2}\otimes I_{2}|\psi\rangle$ can be written as the sum of
$\langle\psi|HH\rangle\langle HH|\psi\rangle$,
$\langle\psi|HV\rangle\langle HV|\psi\rangle$,
$\langle\psi|VH\rangle\langle VH|\psi\rangle$,
$\langle\psi|VV\rangle\langle VV|\psi\rangle$;
$\langle\psi|A_{11}\rangle\langle A_{11}|\otimes I_{2}|\psi\rangle$ can be decomposed into
$\langle\psi|A_{11}H\rangle\langle A_{11}H|\psi\rangle$ and
$\langle\psi|A_{11}V\rangle\langle A_{11}V|\psi\rangle$;
Similarly, $\langle\psi| I_{2}\otimes|B_{11}\rangle\langle B_{11}|\psi\rangle$ can be written as the sum of
$\langle\psi|HB_{11}\rangle\langle HB_{11}|\psi\rangle$ and
$\langle\psi|VB_{11}\rangle\langle VB_{11}|\psi\rangle$.
So we can easily calculate these probability,
\begin{eqnarray}\label{}
\left\{
\begin{array}{l}
\langle\psi|I_{2}\otimes I_{2}|\psi\rangle=\langle\psi|HH\rangle\langle HH|\psi\rangle+\langle\psi|HV\rangle\langle HV|\psi\rangle+\langle\psi|VH\rangle\langle VH|\psi\rangle+\langle\psi|VV\rangle\langle VV|\psi\rangle=N_{\rm total}/N_{\rm total},\\
\langle\psi|A_{11}\rangle\langle A_{11}|\otimes I_{2}|\psi\rangle=\langle\psi|A_{11}H\rangle\langle A_{11}H|\psi\rangle+\langle\psi|A_{11}V\rangle\langle A_{11}V|\psi\rangle=(N_{9}+N_{10})/N_{\rm total},\\
\langle\psi| I_{2}\otimes|B_{11}\rangle\langle B_{11}|\psi\rangle=\langle\psi|HB_{11}\rangle\langle HB_{11}|\psi\rangle+\langle\psi|VB_{11}\rangle\langle VB_{11}|\psi\rangle=(N_{13}+N_{14})/N_{\rm total},\\
\langle\psi|A_{11}\rangle\langle A_{11}|\otimes|B_{11}\rangle\langle B_{11}|\psi\rangle=N_{5}/N_{\rm total},
\end{array}
\right.
\end{eqnarray}
where $N_{\rm total}=\sum_{i=1}^{i=4}N_{i}$. Here, we get the the expression of $P_{8}$,

\begin{eqnarray}
P_{8}&=&\langle\psi|I_{2}\otimes I_{2}|\psi\rangle-\langle\psi|A_{11}\rangle\langle A_{11}|\otimes I_{2}|\psi\rangle-\langle\psi| I_{2}\otimes|B_{11}\rangle\langle B_{11}|\psi\rangle+\langle\psi|A_{11}\rangle\langle A_{11}|\otimes|B_{11}\rangle\langle B_{11}|\psi\rangle,\nonumber\\
&=&(N_{\rm total}-N_{9}-N_{10}-N_{13}-N_{14}+N_{5})/N_{\rm total}.
\end{eqnarray}

In the same way, we get other probabilities $P_{1-7}$.

\begin{eqnarray}\label{P_exp}
\left\{
\begin{array}{llllllll}
P_{1} & = & N_{5}/N_{\rm total},\\
P_{2} & = & (N_{15}+N_{16}-N_{8})/N_{\rm total},\\
P_{3} & = & (N_{\rm total}-N_{11}-N_{12}-N_{13}-N_{14}+N_{7})/N_{\rm total},\\
P_{4} & = & N_{6}/N_{\rm total},\\
P_{5} & = & N_{7}/N_{\rm total},\\
P_{6} & = & (N_{\rm total}-N_{9}-N_{10}-N_{15}-N_{16}+N_{6})/N_{\rm total},\\
P_{7} & = & (N_{11}+N_{12}-N_{8})/N_{\rm total}.\\
%P_{8} & = & (N_{\rm total}-N_{9}-N_{10}-N_{13}-N_{14}+N_{5})/N_{\rm total}\\
\end{array}
\right.
\end{eqnarray}

\subsection{The experimental result of general Hardy-type paradoxes based on the Bell inequality $\mathcal{I}_{4422}\leq 10$}

\begin{figure}[th]
	\begin{center}
		\includegraphics[width=80mm]{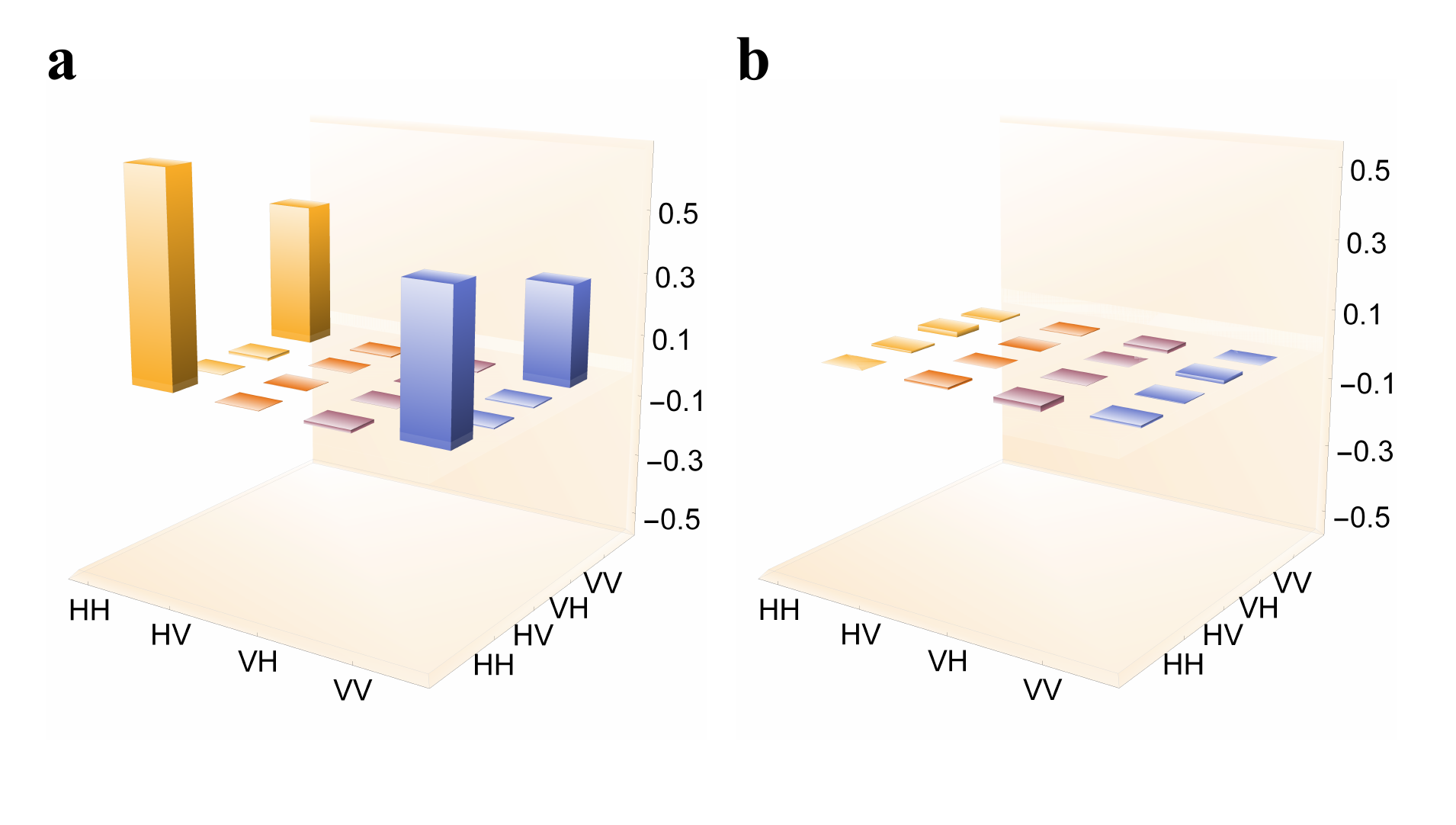}
		\caption{The real part \textbf{a.} and imaginary part \textbf{b.} of the prepared two-qubit density matrix for verifying the Hardy-type paradox based on the Bell inequality $\mathcal{I}_{4422}\leq 10$.}	
		\label{tomo_P26}
	\end{center}
\end{figure}

To show the max conflict of the general Hardy-type paradox based on the Bell inequality $\mathcal{I}_{4422}\leq 10$, we set $\theta= 0.595$. The tomography result is shown in Fig. \ref{tomo_P26}, and the fidelity is $0.99\pm0.05$. Similar to the Hardy-type paradox based on the Bell inequality $\mathcal{I}_{2222}\leq 3$ (or $\mathcal{I}_{\rm CHSH}\leq 3$), here we record the coincident counts in 36 different settings, which are shown in Table \ref{Proj}.

\begin{table}[htbp]
	\caption{ Result of Projective Measurement\tnote{1}}
	\begin{ruledtabular}
		\begin{tabular}{cccccccccc}		
			Alice, Bob  & Coincidence  & Alice, Bob & Coincidence & Alice, Bob  & Coincidence  & Alice, Bob & Coincidence & Alice, Bob & Coincidence\\
			Settings & Counts  & Settings & Counts  & Settings & Counts & Settings & Counts  & Settings & Counts \\
			\hline
			$HH$ & $N_{1}=7166$ & $A_{11}B_{21}$ & $N_{9}=2688$ & $A_{11}B_{41}$ & $N_{17}=14$ & $A_{31}H$ & $N_{25}=1806$ & $HB_{31}$ & $N_{33}=3796$ \\
			$HV$ & $N_{2}=18$ & $A_{21}B_{21}$ & $N_{10}=4432$ & $A_{21}B_{41}$ & $N_{18}=2885$ & $A_{31}V$ & $N_{26}=2737$ & $VB_{31}$ & $N_{34}=1649$\\
			$VH$ & $N_{3}=10$ & $A_{31}B_{21}$ & $N_{11}=252$ & $A_{31}B_{41}$ & $N_{19}=749$ & $A_{41}H$ & $N_{27}=2175$ & $HB_{41}$ & $N_{35}=175$\\
			$VV$ & $N_{4}=3093$ & $A_{41}B_{21}$ & $N_{12}=148$ & $A_{41}B_{41}$ & $N_{20}=3153$ & $A_{41}V$ & $N_{28}=2538$ & $VB_{41}$ & $N_{36}=3159$\\
			$A_{11}B_{11}$ & $N_{5}=3193$ & $A_{11}B_{31}$ & $N_{13}=2158$ & $A_{11}H$ & $N_{21}=3137$ & $HB_{11}$ & $N_{29}=28$&&\\
			$A_{21}B_{11}$ & $N_{6}=1167$ & $A_{21}B_{31}$ & $N_{14}=5099$ & $A_{11}V$ & $N_{22}=3320$ & $VB_{11}$ & $N_{30}=3045$&&\\
			$A_{31}B_{11}$ & $N_{7}=2839$ & $A_{31}B_{31}$ & $N_{15}=57$ & $A_{21}H$ & $N_{23}=3907$ & $HB_{23}$ & $N_{31}=3202$&&\\
			$A_{41}B_{11}$ & $N_{8}=2559$ & $A_{41}B_{31}$ & $N_{16}=22$ & $A_{11}V$ & $N_{24}=1265$ & $VB_{24}$ & $N_{32}=1834$&&\\	
			
		\end{tabular}\label{Proj}
	\end{ruledtabular}
	\begin{tablenotes}
		\footnotesize
		\item[1] $|\psi\rangle= \cos\theta|HH\rangle+\sin\theta|VV\rangle$; $A_{i1}=\cos\theta_{ai}\vert H\rangle+\sin\theta_{ai}\vert V\rangle$;
		$B_{i1}= \cos\theta_{bi}\vert H\rangle+\sin\theta_{bi}\vert V\rangle$;\\
		$\theta=0.595,\;\;\theta_{a1}=1.430,\;\; \theta_{a2}=0.710,\;\; \theta_{a3}=2.050,\;\; \theta_{a4}=2.114,\;\; \theta_{b1}=1.641,\;\; \theta_{b2}=0.958,\;\; \theta_{b1}=0.808,\;\; \theta_{b2}=-1.380$.
	\end{tablenotes}
\end{table}

% The coincidence window is 3ns, and the \rm total number of photons collected within the integral time 1 second.
According to the similar method shown above, $P_{1}-P_{26}$ can be obtained as follows:
\begin{eqnarray}\label{P_exp2}
\left\{
\begin{array}{llllllll}
P_{1} & = & (N_{\rm total}-N_{21}-N_{22}-N_{31}-N_{32}+N_{9})/N_{\rm total},\\
P_{2} & = & N_{14}/N_{\rm total},\\
P_{3} & = & N_{13}/N_{\rm total},\\
P_{4} & = & 2\times(N_{25}+N_{26}-N_{15})/N_{\rm total},\\
P_{5} & = & N_{10}/N_{\rm total},\\
P_{6} & = & (N_{\rm total}-N_{25}-N_{26}-N_{29}-N_{30}+N_{7})/N_{\rm total},\\
P_{7} & = & (N_{23}+N_{24}-N_{18})/N_{\rm total},\\
P_{8} & = & (N_{\rm total}-N_{27}-N_{28}-N_{29}-N_{30}+N_{8})/N_{\rm total},\\
P_{9} & = & N_{11}/N_{\rm total},\\
P_{10} & = & 2\times(N_{33}+N_{34}-N_{15})/N_{\rm total},\\
P_{11} & = & (N_{\rm total}-N_{21}-N_{22}-N_{35}-N_{36}+N_{17})/N_{\rm total},\\
P_{12} & = & N_{17}/N_{\rm total},\\
P_{13} & = & N_{6}/N_{\rm total},\\
P_{14} & = & (N_{\rm total}-N_{23}-N_{24}-N_{29}-N_{30}+N_{6})/N_{\rm total},\\
P_{15} & = & (N_{\rm total}-N_{23}-N_{24}-N_{33}-N_{34}+N_{14})/N_{\rm total},\\
P_{16} & = & N_{7}/N_{\rm total},\\
P_{17} & = & (N_{\rm total}-N_{23}-N_{24}-N_{31}-N_{32}+N_{10})/N_{\rm total},\\
P_{18} & = & (N_{35}+N_{36}-N_{18})/N_{\rm total},\\
P_{19} & = & N_{8}/N_{\rm total},\\
P_{20} & = & N_{5}/N_{\rm total},\\
P_{21} & = & N_{9}/N_{\rm total},\\
P_{22} & = & (N_{\rm total}-N_{21}-N_{22}-N_{33}-N_{34}+N_{13})/N_{\rm total},\\
P_{23} & = & (N_{\rm total}-N_{25}-N_{26}-N_{31}-N_{32}+N_{11})/N_{\rm total},\\
P_{24} & = & (N_{31}+N_{32}-N_{12})/N_{\rm total},\\
P_{25} & = & (N_{27}+N_{28}-N_{12})/N_{\rm total},\\
P_{26} & = & (N_{\rm total}-N_{21}-N_{22}-N_{29}-N_{30}+N_{5})/N_{\rm total}.\\
\end{array}
\right.
\end{eqnarray}

\begin{figure}[th]
	\begin{center}
		\includegraphics[width=150mm]{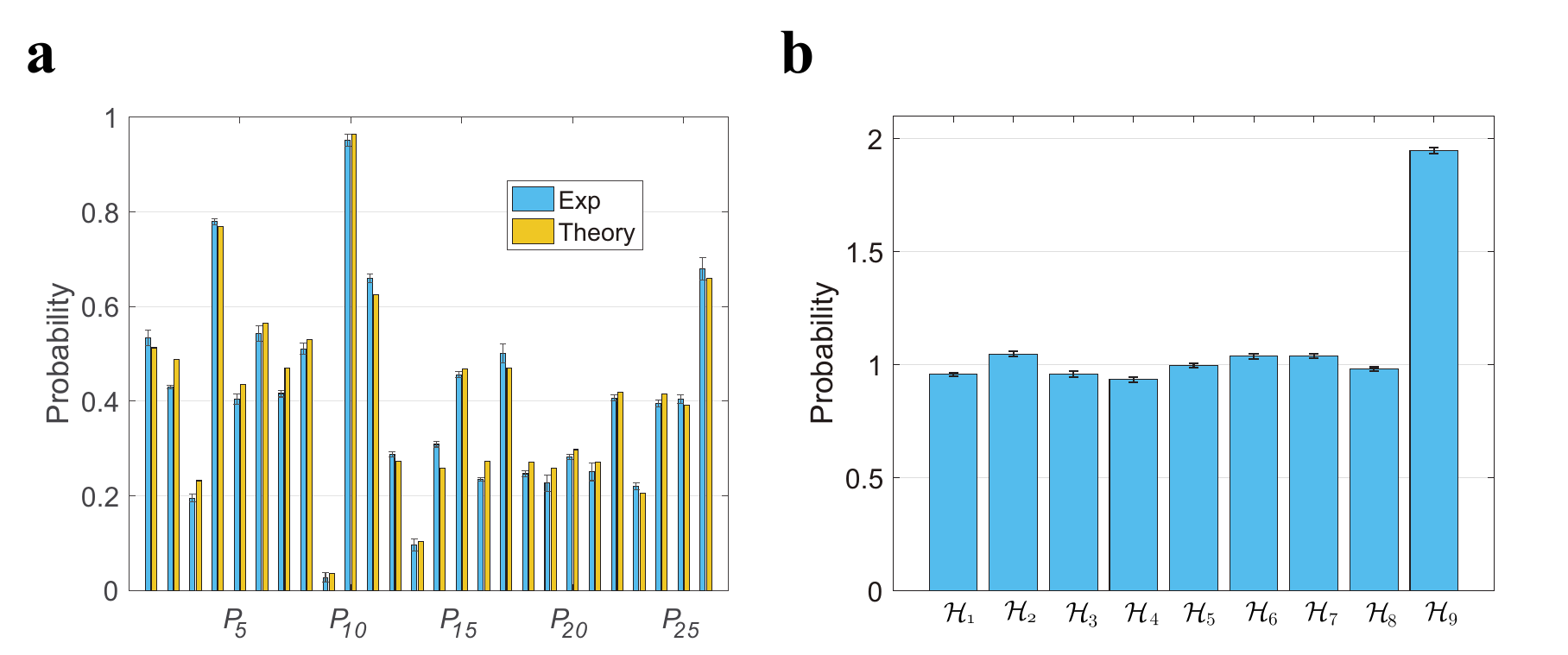}
		\caption{Experimental results. {\bf a.} The probabilities of $P_{1}$ to $P_{26}$. Blue and yellow columns represent the experimental and theoretical results, respectively. ${\bf b.}$ The experimental values of the 9 Hardy's constraints, which coincide with the corresponding theoretical predictions shown in (\ref{eqI4422}).
		%Error bars are deduced from the counting statistics.
		}	
		\label{P26}
	\end{center}
\end{figure}

The probabilities are shown in Fig. \ref{P26}a. The blue columns are the experimental results and the yellow columns represent the theoretical predictions. The experimental results agree well with the theoretical predictions. The conflict of the Hardy's paradox can be reached as high as $P_{26}=0.6802\pm0.0238$, which is about seven times larger than the original one ($\approx 9\%$)~\cite{hardy93}. We further show the 9 Hardy's constraints of $\mathcal{H}_{i}\;  (i=1,2\cdots9)$ obtained in experiment in Fig. \ref{P26}b, where $\mathcal{H}_{1}\sim \mathcal{H}_{8}$ are nearly equal to 1 and $\mathcal{H}_9$ is nearly equal to 2. All the 9 Hardy's constraints are well confirmed within the experimental errors. Error bars are deduced from the counting statistics, which are assumed to be Poissonian distribution.

%\begin{figure}[th]
%	\begin{center}
%		\includegraphics[width=0.8\columnwidth]{tomo_P26}
%		\caption{The real part \textbf{a.} and imaginary part \textbf{b.} of the prepared two-qubit density matrix for verifying the Hardy-type paradoxes based on the Bell inequality $\mathcal{I}_{4422}\leq 10$.}	
%		\label{tomo_P26}
%	\end{center}
%\end{figure}

